%% file: vis.tex
\documentclass[5p,authoryear]{elsarticle}
\usepackage{mathtools}
\usepackage{amsmath}
\usepackage{graphicx}
\usepackage{times}
\usepackage{amssymb}
\usepackage{multirow}
\usepackage{algorithm}
\usepackage{algorithmicx}
\usepackage{algpseudocode}
\usepackage{xcolor}
\usepackage{textcomp}
\usepackage{eqparbox}
\usepackage{hyperref}
\usepackage{wrapfig}
\usepackage{subfig}
\usepackage{tabularx}
\usepackage{subfiles}
\usepackage[explicit]{titlesec}

\titleformat{\section}{\normalfont}{\textbf{\thesection}}{1em}{\MakeUppercase{\textbf{#1}}}
\titlespacing*{\section}{0em}{2em}{0.8em}
\titleformat{\subsection}{\normalfont}{\textbf{\thesubsection}}{1em}{\textbf{#1}}
\titlespacing*{\subsection}{0em}{2em}{0.8em}
\titleformat{\subsubsection}{\normalfont}{\textit{\thesubsubsection}}{1em}{\textit{#1}}
\titlespacing*{\subsubsection}{0em}{2em}{0.8em}

\graphicspath{{figs/}{../figs/}}
\usepackage{textcomp}
\usepackage{float}

\begin{document}

\title{Vizic: A Jupyter-based Interactive Visualization Tool for Astronomical Catalogs
}

\author[phy,ncsa]{Weixiang Yu \corref{email1}}
\author[ncsa,astro]{Matias Carrasco Kind}
\author[astro,ncsa]{Robert J. Brunner}
\address[phy]{Department of Physics, University of Illinois at Urbana-Champaign, Urbana IL, USA}
\address[ncsa]{National Center for Supercomputing Applications, University of Illinois at Urbana-Champaign, Urbana IL, USA}
\address[astro]{Department of Astronomy, University of Illinois at Urbana-Champaign, Urbana IL, USA}
\cortext[email1]{wyu16@illinois.edu}
\date{outline}

\begin{abstract}
The ever-growing datasets in observational astronomy have challenged scientists in many aspects, including an efficient and interactive data exploration and visualization.
Many tools have been developed to confront this challenge. However, they usually focus on displaying the actual images or focus on visualizing patterns within catalogs in a predefined way.
In this paper we introduce \textit{Vizic}, a Python visualization library that builds the connection between images and catalogs through an interactive map of the sky region.
\textit{Vizic} visualizes catalog data over a custom background canvas using the shape, size and orientation of each object in the catalog.
The displayed objects in the map are highly interactive and customizable comparing to those in the observation images.
These objects can be filtered by or colored by their property values, such as redshift and magnitude. They also can be sub-selected using a lasso-like tool for further analysis using standard Python functions and everything is done from inside a Jupyter notebook. Furthermore, \textit{Vizic} allows custom overlays to be appended dynamically on top of the sky map.
We have initially implemented several overlays, namely, Voronoi, Delaunay, Minimum Spanning Tree and HEALPix grid layer, which are helpful for visualizing large-scale structure. All these overlays can be generated, added or removed interactively with just one line of code.
The catalog data is stored in a non-relational database, and the interfaces have been developed in JavaScript and Python to work within Jupyter Notebook, which allows to create customizable widgets, user generated scripts to analyze and plot the data selected/displayed in the interactive map. This unique design makes \textit{Vizic} a very powerful and flexible interactive analysis tool.
\textit{Vizic} can be adopted in variety of exercises, for example, data inspection, clustering analysis, galaxy alignment studies, outlier identification or just large scale visualizations.
\end{abstract}


\maketitle

\section{Introduction}
For the past decades, conducting large-area sky surveys became one of the most powerful approaches to carry out observations within the astronomy community in order to understand the Universe. The Sloan Digital Sky Survey (SDSS) \citep{sdssYork2000} set the foundations for such surveys followed by others, such as DES and Pan-STARRS \citep{des, PanSTARRS}. Modern telescopes and world-class supercomputing facilities produced exceptionally good images and exceptionally precise measurements of astrophysical sources, challenging astronomers to face unprecedentedly massive datasets.
Moreover, future large sky, for example, LSST and Euclid \citep{lsstOverview,lsstSci,euclid}, will obtain data of orders of magnitude larger than what we have today. These massive datasets place a new challenge on astronomers in the light of data exploration and visualization.

Various groups and individuals have tried to confront this challenge by developing new astronomical applications that are capable of displaying and visualizing large datasets. Notable examples include \textit{Aladin Lite} \citep{AladinLite2014}, \textit{VisiOmatic} \citep{Bertin2015}, \textit{ASCOT} and \textit{Toyz} \citep{Moolekamp2015}, to mention just a few.
The majority of these softwares are web-based applications that focus on the visualization of high-resolution images with additional functionalities of catalog over-plotting or image analysis. Meanwhile, \textit{GLUE} \citep{GlueViz}, a desktop application that specializes in visualizing selection propagations between different plots of the same dataset, has became a popular tool for exploring hidden patterns within catalogs and images.
Unfortunately all these mentioned tools, while being very powerful, have some drawbacks when it comes to the study of large-scale structure and flexible customization.
Image-focused applications are limited by the fundamental characters of photometric observations, whereas images produced by sky surveys are fixed in terms of compositions, in other words, the sources are immoveable. In this regard, catalogs are more flexible considering astronomical sources in a catalog can be easily sorted and filtered.
However, tools that are more friendly with tabular data are usually less good at interactive visualizations of geospatial data.
They either render data points into simple scatter plots or summarize data distribution into pixels like with \textit{VaeX} \citep{vaex}, which lacks rich interactions with individual object while providing a very powerful way to visualize distributions of extremely large catalogs.
The regarding limitations of existing softwares in addition to emerging new technologies for building interactive web tools have generated a need to improve existing tools and develop new ones.

Thanks to the increasingly powerful web technologies (e.g., HTML5 \citep{HTML5}, CSS \citep{CSS} and JavaScript \citep{JS}) and the improved data bandwidth, nowadays much more complex tasks can be carried out in a browser.
Along with the great availability of graphic design JavaScript libraries, web browser has became a fertile ground for visualization projects. In addition, thanks to projects like Jupyter \citep{Jupyter} we now have the powerful combination of a web-based application that can run anywhere with the flexibility of a Python scripting environment where scientist can customize plots, algorithms and workflows only restricted by the limitations of the programming language.

Considering astronomers are usually experienced scripting programmers (but less so in web development) and the fact that Python is one of the most common languages used today, we have created \textit{Vizic}, a Jupyter-based interactive visualization tool, which is a Python package designed to work with the Jupyter Notebook App.
This tool (formally IPython \citep{ipython} Notebook) is a server-client application that allows creating and running Jupyter notebooks in which Python code can be executed interactively in executable cells. Unlike other existing tools, \textit{Vizic} tries to make the connection between a source catalog and its images by drawing astronomical sources on an interactive map, where we can clearly visualize the spatial distribution of the observed objects, and at the same time keeping full control over the compositions, through objects filtering, color mapping, over plotting layers, among other useful features to visualize patterns on the data in a much more efficient manner.
Since Jupyter Notebook App is a server-client system, \textit{Vizic} can be used for accessing data archives at remote locations as well as working with catalogs stored on a local machine. Regarding the rising popularity of Jupyter Notebook App within the science community, we believe in providing an interactive interface without limiting the scripting ability to extend and to use this package, which will enable a faster scientific discovery.

The rest of the article is organized as follows. In Section \ref{package}, we start with a brief introduction to \textit{Vizic} and highlight some of its important features.
In Section \ref{technical}, we discuss the technical details regarding the architecture and dependencies of \textit{Vizic}.
In Section \ref{application}, we present an example application of \textit{Vizic} in visualizing large-scale structure.
Section \ref{install} provides the installation instruction, and Section \ref{performance} assesses the performance of \textit{Vizic} with various loads.
Lastly, in Section \ref{future}, we gives the conclusions, and describes future plans and possible extensions.

\section{Vizic}
\label{package}
\begin{figure}[h]
  \centering
  \includegraphics[width=0.45\textwidth]{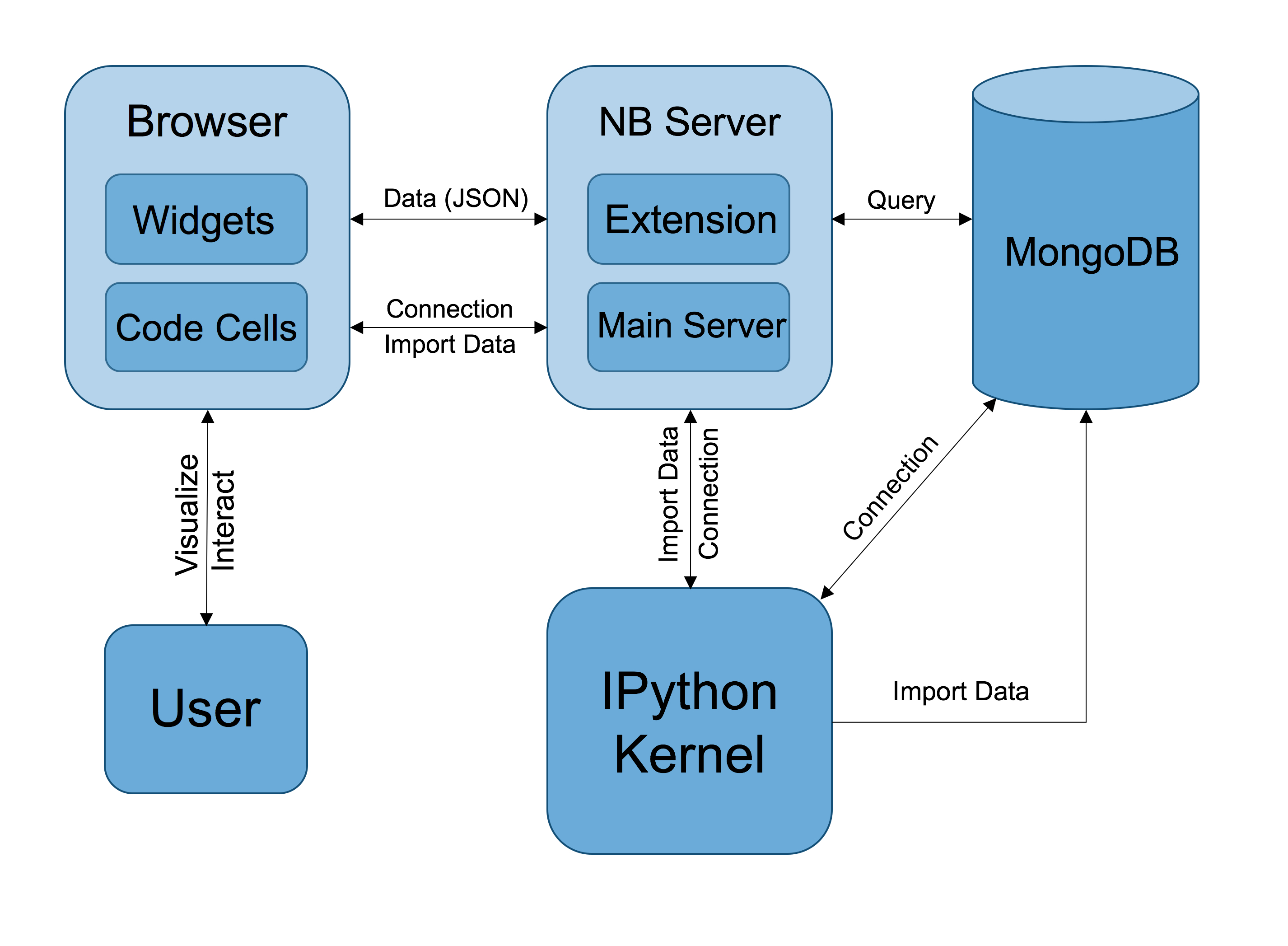}
  \caption{A diagram showing the workflow of \textit{Vizic}. ``NB Server'' refers to Jupyter Notebook server.}
  \label{fig:model_d}
\end{figure}

\textit{Vizic} is a Python and Javascript library, which provides the API to display and interact with sky maps created from catalogs within Jupyter notebooks.
At the front-end, the interactive sky maps are generated using \textit{Leaflet}\footnote{http://leafletjs.com/} \citep{leaflet}, a popular JavaScript library for building interactive map applications on the web, and rendered through the widget framework provided by \textit{ipywidgets}\footnote{https://ipywidgets.readthedocs.io/en/latest/} \citep{ipywidgets}, which is a library of interactive HTML widgets for Jupyter notebooks and IPython kernels.
\textit{Vizic} works side by side with MongoDB\footnote{https://www.mongodb.com/} \citep{mongo}, therefore an active MongoDB instance, either running in the background or in a remote location (see Figure \ref{fig:model_d} for a workflow diagram), is required when \textit{Vizic} is used.
MongoDB is a non SQL, document-oriented database, which is built for scalability and performance. Since the data is stored in documents instead of multiple tables with complex relations, MongoDB database can spread the data over different servers and nodes. In the case of storing simple astronomical catalogs, MongoDB is particularly suitable. MongoDB also provides a geospatial index for making extremely efficient location-based queries. The rational to use MongoDB is discussed in more detail in Section \ref{mongodb}.

The interactive sky maps can be easily created from catalogs stored in pandas \citep{pandas} DataFrames or catalogs that are previously ingested into the database by providing the collection names. Data stored in other formats, for example, \textit{Astropy} \citep{astropy} Table object, can also be fed into \textit{Vizic} after a simple conversion to pandas DataFrame.
A \textit{pandas} DataFrame is a Python object that handles arrays and tabular data very efficiently.\footnote{http://pandas.pydata.org}
When a DataFrame is provided, \textit{Vizic} will format the data to match the mapping mechanism used, which is described in more detail in Section \ref{slippy}.
In the sky maps, astronomical objects are drawn using their shapes, sizes and rotation angles determined by source catalog extractor softwares such as \textit{SExtractor} \citep{sextractor} to cite an example.
If relevant shape information is not available, the objects will be drawn as filled circles with appropriate radius provided in the catalog. If neither the shape data nor the size data is provided, the objects will be drawn as filled circles with a fixed radius of $0.5$ pixel at the minimum zoom level, however alternative sizes can be specified as an argument, and scaled up as the map is zoomed in.

There are two main categories of widgets provided by \textit{Vizic}, namely, map widgets and control widgets. Map widgets are responsible for creating and managing sky maps and overlays. The core map widgets include \texttt{AstroMap} and \texttt{GridLayer}. The \texttt{AstroMap} widget is the container for all map layers, and the \texttt{GridLayer} widget is the tile-based layer for displaying the astronomical sources or simply the catalog layer. Additional custom overlay widgets are also provided for aid in visualizing large-scale structure (see Section \ref{adv}).
The control widgets are the main interfaces that take advantage of the extensive interactivity of the displayed objects (see Section \ref{interaction}).
For example, through control widgets, we can filter the displayed objects by their property values and apply custom colormaps to these sources. These control widgets do not possess mutual reliance, therefore they can be used independently from each other.
Such a modularized design brings a much cleaner graphical user interface (GUI) comparing to other tools. Users can create their own application and interface by selecting only the needed components and interact with the data and the maps from within the Jupyter notebooks.
After all, \textit{Vizic} provides the foundation for a complete and customizable analysis, which is complementary to the standard scripting analysis workflows.

\subsection{Sky Maps}
\label{basic}
Before any sky maps can be created, the connection to a running MongoDB instance has to be established.
The connection could be initiated by creating a \texttt{Connection} class object with a URL that specifies the address of the MongoDB instance and a port number. In the background, the IPython kernel creates a MongoDB client using the provided information and assigns the client to an attribute of the \texttt{Connection} object, which is required for creating \texttt{GridLayer} widgets.
After the connection is successfully established, the \texttt{AstroMap} widget and the \texttt{GridLayer} widget instances can be created by initializing the corresponding Python object. To display catalogs in the notebook cells, we need to add the \texttt{GridLayer} widget to the \texttt{AstroMap} widget using the \texttt{add\_layer} function.
Then we can zoom and pan the map interactively as with a Google Map\footnote{https://www.google.com/maps}
The transformation of the celestial coordinates is accomplished by a customized coordinate reference system (CRS) JavaScript library that determines the projection scale based on the extent of the map measured in degrees. In this manner, we can use RA and DEC information from the sources directly.

When creating catalog layers from \textit{pandas} DataFrames, the default columns used for reading the locations are the standard RA and DEC. The default columns used for reading the shape and orientation of each object are A\_IMAGE, B\_IMAGE, THETA\_IMAGE, which are respectively the semi-major axis, semi-minor axis and the rotation angle used in the Scalable Vector Graphics (SVG) \citep{SVG} ellipse adopted to represent that object. This process is repeated for every object in the catalog.
Different column names can also be specified in the arguments for reading this information.
\begin{figure*}[h!]
	\centering
	\includegraphics[width=\textwidth]{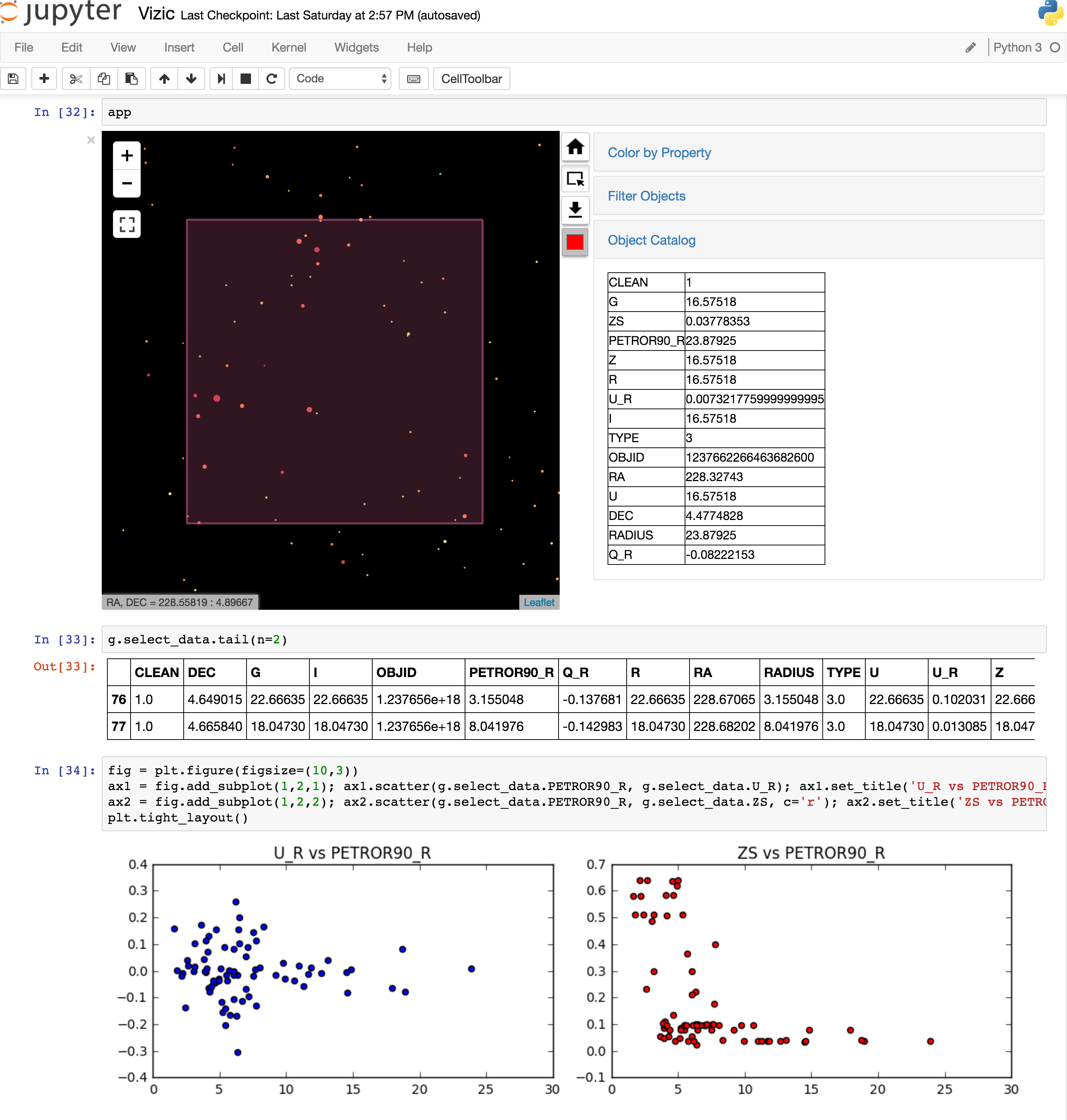}
	\caption{A GUI created from widgets provided by \textit{Vizic}, where the dashboard is completely customizable. The pink rectangle on the sky map is the selection bounding box. The table below the ``Object Catalog'' toolbar displays the catalog entry for a clicked object. The middle section shows the data returned by the selection tool in a \textit{pandas} DataFrame for direct manipulation through the notebook.
	Two plots at the bottom of this figure are generated using the data returned by the selection tool.
    The objects on the sky map are colored by their magnitude in $I$ band. And the size of these objects are scaled up by a factor of 3 to better visualize the effect of color mapping. }
	\label{fig:overall}
\end{figure*}

\subsection{Interaction}
\label{interaction}
\subfile{sections/interaction}

\subsection{Custom Overlay}
\label{adv}
\subfile{sections/advanced}

\section{Architecture \& Dependencies}
\label{technical}

\subfile{sections/architecture}

\section{Example Application}
\label{application}
\begin{figure}[h]
    \centering
    \subfloat[The trimmed MST with $r_{max}=0.05^\circ$ (0.6 Mpc/h at z=0.2) and $n_{min}=30$]{\includegraphics[width=0.4\textwidth]{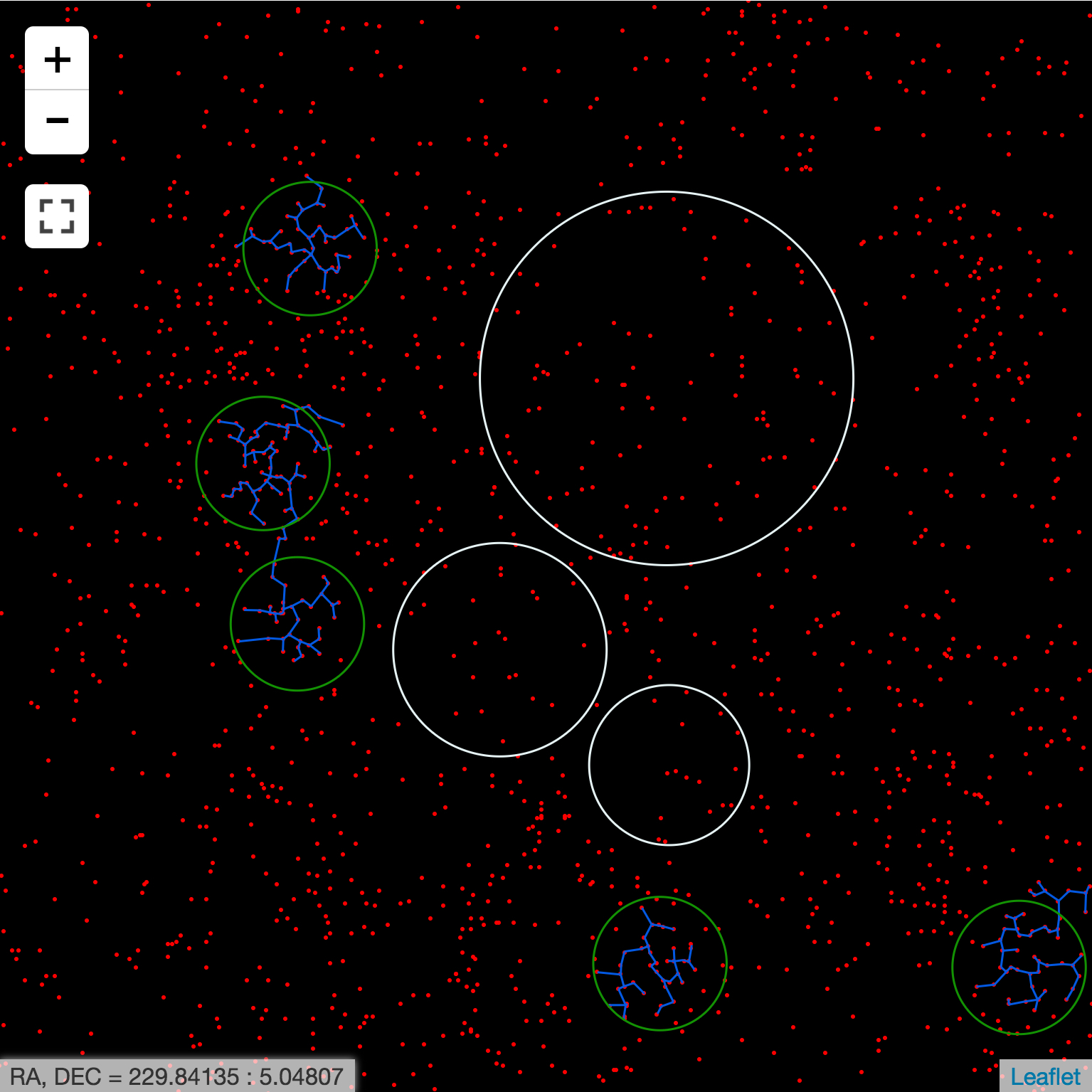}\label{fig:app1}}
    \hfill
    \subfloat[The trimmed MST with $r_{max}=0.08^\circ$ ((0.957 Mpc/h at z=0.2)) and $n_{min}=20$]{\includegraphics[width=0.4\textwidth]{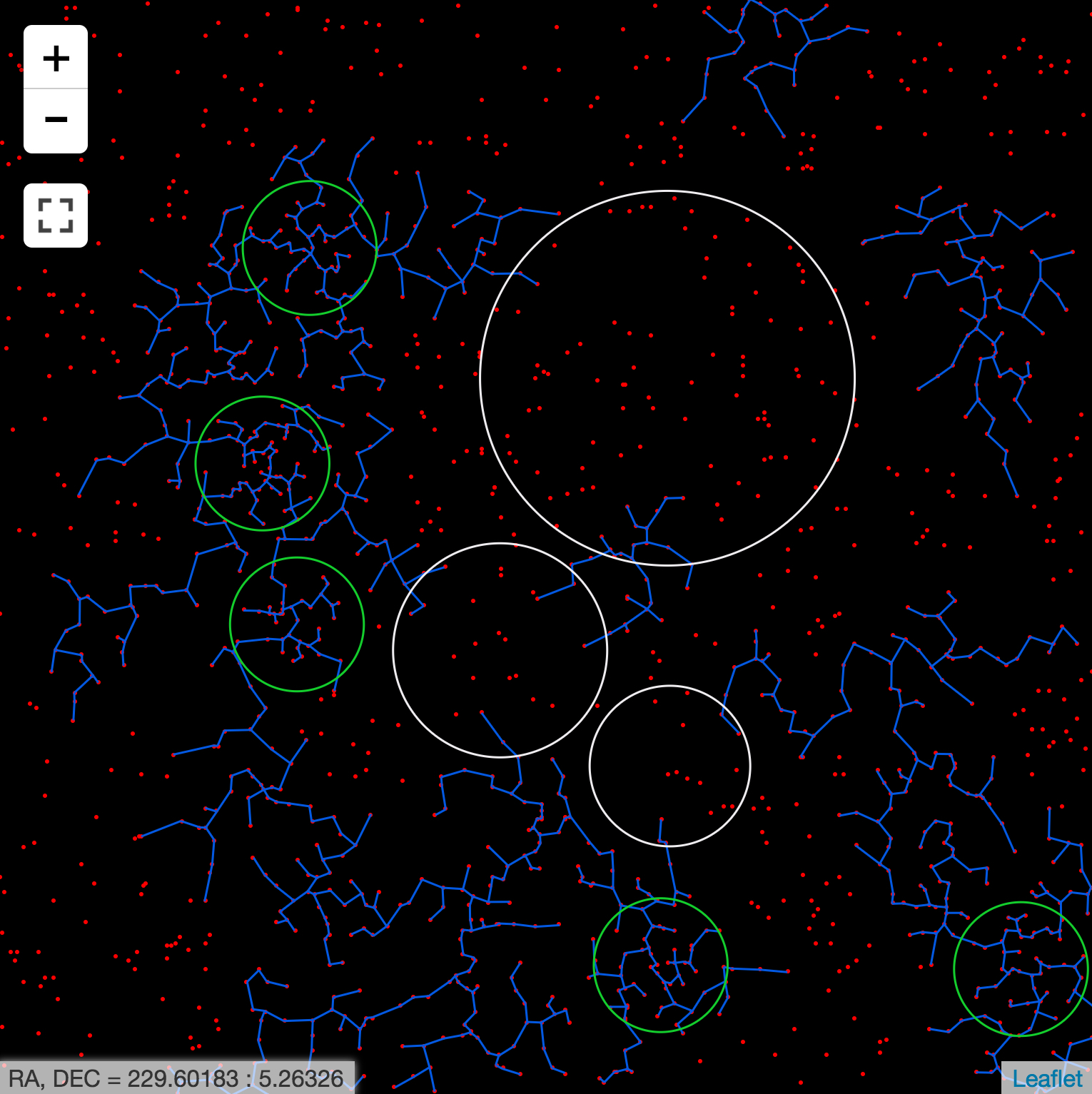}\label{fig:app2}}
    \caption{Two figures showing different combination of $r_{max}$ and $n_{min}$ to prune the full MST to identify structures. The green circles mark the galaxy clusters while the white circles mark the voids.}
    \label{fig:app}
\end{figure}
In this section, we present an example usage of \textit{Vizic} in scientific research. We first create an interactive map using a subset of the spectroscopic catalog from SDSS Data Release 13 (DR13) \citep{dr13, sdss_cam}.
This subset covers the area from $227^\circ$ to $230^\circ$ in right ascension and from $4^\circ$ to $7^\circ$ in declination with a total of nine square degrees.
This sample catalog contains only the objects from the $[0, 0.2]$ redshift bin \citep{sdss_spec}. For demonstration purpose, we choose not to provide the shape and size information when creating the sky map.

Then we look for galaxy clusters and voids on the sky map and mark them with circle layers using different colors. For the sake of this example, the positions and sizes are selected by eye.
The approximated centers of the clusters and voids are determined by the mouse position tracker located on the bottom left corner of the map (See Figure \ref{fig:app}). Next, we apply the MST overlay onto the sky map.
By pruning the tree with different combination of $r_{max}$ and $n_{min}$, we can confirm our previous structure selections with the tree groups that are kept.
We can observe that for a more strict cut in $r_{max}$, i.e. reducing maximum linkage length between two points, as shown in Figure \ref{fig:app1}, compact structures selected by the MST are correlated to those circles shown in green (i.e., clusters), while for a more relaxed cut, in Figure \ref{fig:app2}, the structures shown follow the underlying filament structure around voids (white circles) much better.

Additionally, other layers can be over-plotted, data can be selected for further inspection, objects can be filtered, colored and scaled to aid with visualization and analysis.
Multiple maps, multiple cuts and multiple datasets can be used interactively with other scripting  procedures and functions on the data from other cells in one single Jupyter notebook, increasing the potential and effectiveness of \textit{Vizic}.

\section{Distribution \& Installation}
\label{install}
\textit{Vizic} is registered with Python Package Index\footnote{https://pypi.python.org/pypi/vizic}, therefore it can be installed with \textit{pip} for Python 3.
It can also be installed from source code by downloading the GitHub repository\footnote{https://github.com/ywx649999311/vizic}.
\textit{Vizic} requires a running MongoDB instance, the installation and setup instructions can be found from the official MongoDB website\footnote{https://docs.mongodb.com/manual/}.
More detailed instructions on the installation of \textit{Vizic} and its required dependencies as well as tutorials, API documentation and examples can be found in the official \textit{Vizic} documentation\footnote{http://www.wx-yu.com/vizic/index.html}.

Instead of installing the package in a local machine and keeping a MongoDB instance running, the user can also build a Docker\footnote{https://www.docker.com} container from the Dockerfile included in the GitHub repository.
Both the MongoDB database and the Jupyter notebook server run inside the container with only designated ports exposed to the host for easy deployment.
An example dataset and notebooks are included as part of the \textit{Vizic} distribution on GitHub as well.

\section{Performance \& Discussion}
\label{performance}

Visualizing large datasets through a browser has always been a difficult task. Unlike some of the existing tools, the goal of \textit{Vizic} is not to display as many objects as possible at once, but instead to reduce the number of displayed sources without losing important information using the ``slippy map'' implementation described before.

We set up several tests to examine the overall performance of \textit{Vizic} regarding data ingestion and map interaction and deployment. All tests were performed on a Macbook Pro running OS X El Capitan equipped with a 2.5GHz Intel Core i7 (i7-4870HQ) processor, 16Gb of RAM and a SSD.
Both MongoDB instance and the Jupyter App server were running locally. Unless specifically stated, \textit{Vizic} was tested using the browser Chrome.

\begin{figure}[h]
  \centering
  \includegraphics[width=0.5\textwidth]{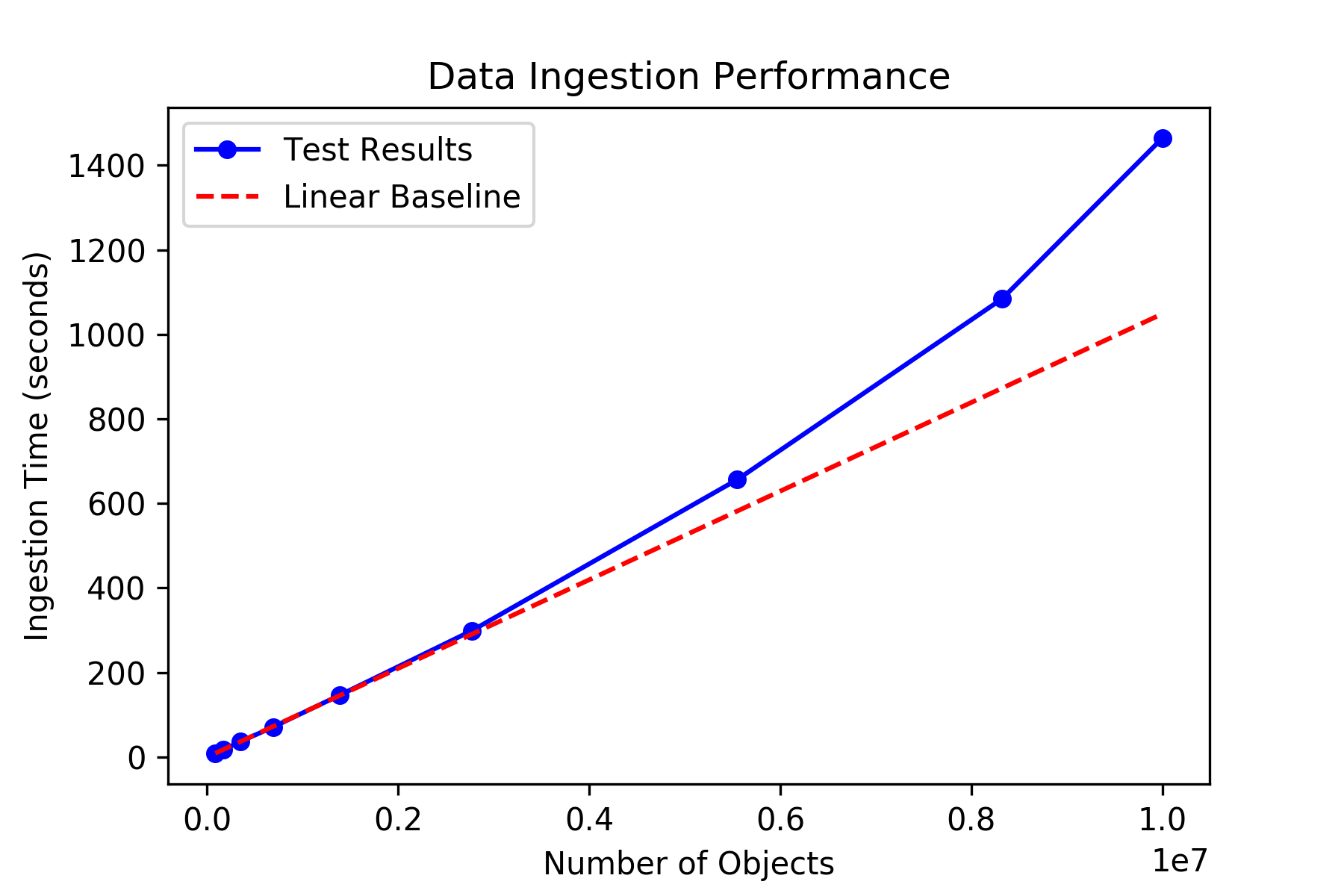}
  \caption{Data ingestion time as a function of catalog size. The linear baseline was created using the first three data points on this plot, showing the diverging character of the performance curve after 3 million objects.}
  \label{fig:ingest}
\end{figure}

Figure \ref{fig:ingest} shows the time in seconds for the data ingestion as a function of catalog size (number of objects), and all catalogs ingested in this test contain fifteen columns. The linear baseline shown in this figure is a trendline for the first three data points on this plot. We notice that the actual performance curve starts to deviate from the linear baseline when the catalog size exceeds 3 million. We presume such diverging behavior as a result of computational resources being exhausted.

\begin{figure}[h]
  \centering
  \includegraphics[width=0.5\textwidth]{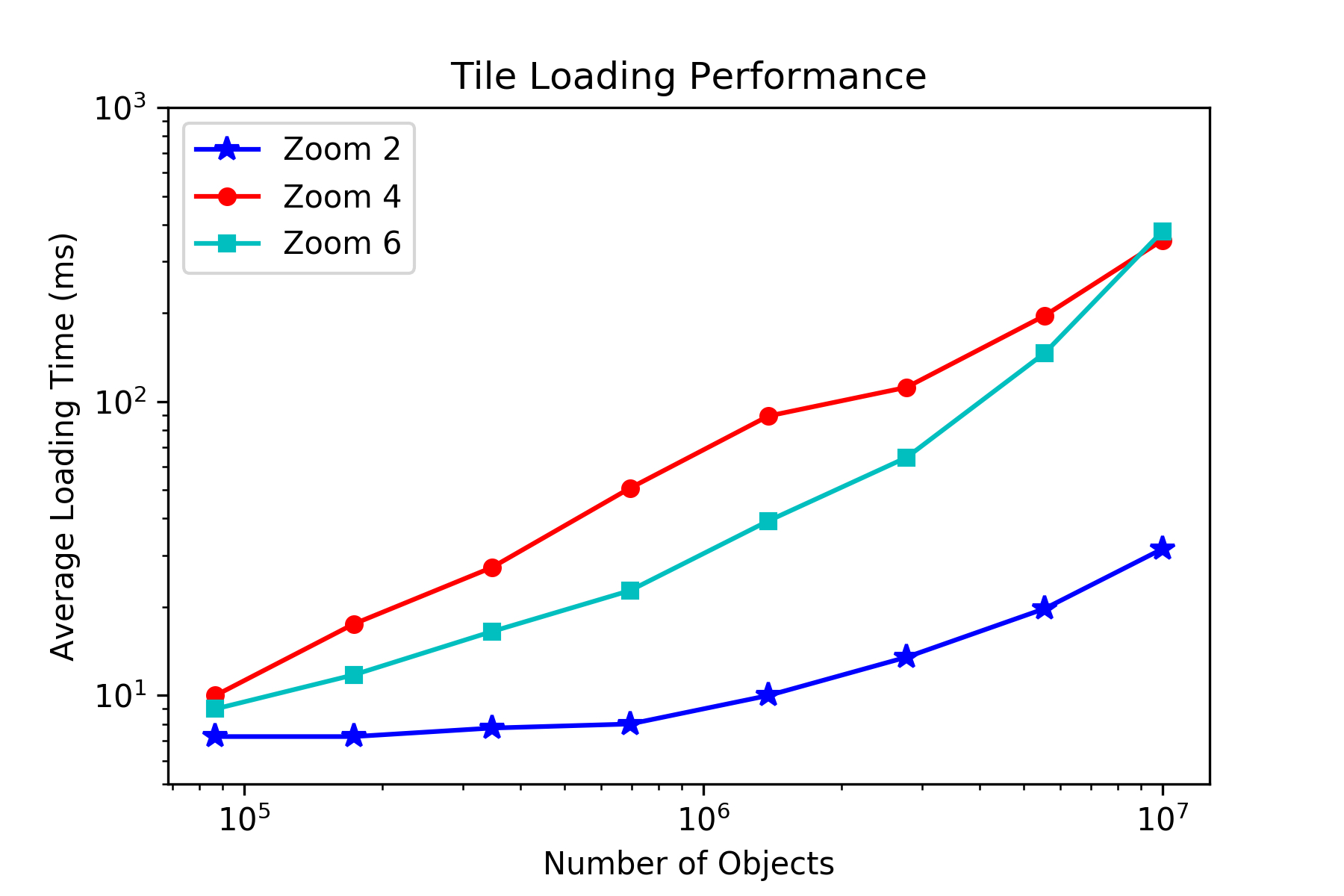}
  \caption{Display response and loading time as a function of catalog size for different zoom levels.}
  \label{fig:load}
\end{figure}

The second test performed, as shown in Figure \ref{fig:load}, profiles the timings for map loading triggered when applying interactively different zoom levels.
To focus on the effect of catalog sizes on the tiles loading performance, we created the test datasets by continuously and repeatedly throwing a collection of objects onto a $15^\circ \times 15^\circ$ area until the desired number of objects is reached, hence increasing the density of the objects each time.
This collection of objects were selected from a same $15^\circ \times 15^\circ$ area in SDSS Data Release 13.

The tile loading time is a direct indication of the responsiveness of the map when zoom level changes interactively.
In our test, we fixed the map window to 512 by 512 pixels and aligned the center of map with the center of the window. Then, four tiles were loaded simultaneously each time the zoom level changes. We used the Chrome DevTool to record the loading time for each tile and take the average as the final result.
Accordingly, if more tiles are loaded at the same time, for example, in the full screen mode, the loading performance might differ but follow a similar trend seen in Figure \ref{fig:load}.

Another critical measurement is the smoothness of the panning motion on the map.
Without triggering new tile loadings, the responsiveness of the panning motion is proportional to the number of SVG elements added to the Document Object Model (DOM) tree, the larger the DOM tree the longer it takes to pan the map.
In the tests performed above, the tile loadings were usually slower than the panning motion and we didn't experience any significant lags, where the maximum number of objects loaded in one tile was around fifteen hundred.
Nonetheless, higher latencies were observed in the full screen mode, as a direct result of increased number of SVG elements added to the DOM tree.

As shown in the performance tests, the largest catalog being visualized on the mentioned machine include 10 million rows and 15 columns. Due to the constraints of the available hardware, we were not able to further examine the limit of \textit{Vizic}. However, we believe \textit{Vizic} will perform well with any dataset that has less than 50 million objects, on a more powerful machine. In the future release, we expect to make \textit{Vizic} capable of handling catalogs on the order of  $10^8$ and above. Currently if a user would like to work with a dataset on that order, we suggest the user to divide the entire dataset into several subareas and then create a map for each subarea. In a case that displaying the whole dataset and observing the big picture is most critical, the users can turn to tools like \textit{VaeX}, but sacrificing the ability to further interact with the data through object filtering and coloring, as well as the application of custom overlays.

At this stage, the pre-defined custom overlays are not suitable for use on a single large dataset (typically over 100K data points being displayed at a time).
However, these overlays could also be made to work by first dividing the region covered by the large catalog into several small areas and then creating a map for each smaller area.
If the interesting objects happen to be near the edges of these small maps, the user can use the provided selection tool to query the catalog from the large map and create another new map from it.
In the future release of \textit{Vizic}, a performance improvement related to the custom overlay is expected.

\section{Conclusions \& Future Work}
\label{future}
In this paper, we present \textit{Vizic}, a Python visualization package, which is designed to work with the Jupyter Notebook App.
To offer an easy and complete analysis environment, \textit{Vizic} utilizes the best part of both images and catalogs by visualizing astronomical sources on an interactive sky map and provides powerful interactions with the Jupyter notebook cells for further interactive scripting analysis with the data.

While resembling the spatial distributions of the sources on the original images, the interactive sky maps created using \textit{Vizic} also allow us to filter or color displayed astronomical objects by their property values.
The lasso-like selection tool provides us the ability to interactively select interesting objects from the sky map and retrieve their catalog from the linked MongoDB database.
Alongside the returned DataFrame containing the catalog for the selected objects, we can further explore and analyze the hidden patterns among the data using various plotting packages available for Python and Jupyter, like \textit{matplotlib}, \textit{scikit-learn} and many others.
In addition, \textit{Vizic} offers us the option to over-plot custom layers on the base map. Four pre-defined custom overlays are included in this package.
The Voronoi diagram overlay, the minimum spanning tree overlay and the Delaunay triangulation overlay provide an easy and fast way to visualize cosmological structures.
The HEALPix grid overlay is a convenient tool to interactively inspect the density field generated by the catalog or to complement with other astronomical data. \textit{Vizic} supports multiple datasets and multiple layers for a full exploration between different catalogs.

We also demonstrate an immediate application of \textit{Vizic} by identifying galaxy clusters, voids and filaments. By utilizing the tree pruning feature provided by the minimum spanning tree overlay, we can match the saved tree branches with previously selected structures to visualize and detect these structures.
The result shows that \textit{Vizic} is a powerful and friendly tool for visualizing large-scale structure. However, its usage is not limited to just galaxy catalogs, any form of catalog data can be easily handled by \textit{Vizic}, such as star and galaxy clusters, star forming regions in a galaxy, etc.

In general, \textit{Vizic} provides a new way to efficiently visualize and interact with astronomical catalogs. Nevertheless, these features provided by \textit{Vizic} can also be applied to data from other scientific fields.

Multiple improvements over the current version has already been planned.
A major improvement is to further increase the scalability of this tool.
While keep increasing the number of objects that \textit{Vizic} can efficiently visualize (with a maximum loading time of 0.5 second for each zoom level), we also would like to match the responsiveness of the custom overlays to that of the tiled base layer.
We are currently testing new method of building these overlay layers using Web Components.
With Web Components we can isolate each overlay from the outside world, so that the CSS style of the elements inside the overlay component is not affected by style changes originated from outside the overlay.
Such feature can dramatically reduce the amount of time consumed by style recalculations when the map shifts locations.

Another improvement is integrating \textit{Vizic} with other interactive plotting libraries and making it even easier for astronomers to explore the catalogs by taking advantage of the scripting power of Python, in a way that multiple tools are connected and the change in one is reflected in the other one, providing a complete user analysis experience in a Jupyter notebook.

\section*{Acknowledgements}
RJB acknowledges support from the National Science Foundation Grant No. AST-1313415.

Some of the results in this paper have been derived using the HEALPix \citep{healpix} package.

Funding for the Sloan Digital Sky Survey IV has been provided by
the Alfred P. Sloan Foundation, the U.S. Department of Energy Office of
Science, and the Participating Institutions. SDSS-IV acknowledges
support and resources from the Center for High-Performance Computing at
the University of Utah. The SDSS web site is www.sdss.org.

SDSS-IV is managed by the Astrophysical Research Consortium for the
Participating Institutions of the SDSS Collaboration including the
Brazilian Participation Group, the Carnegie Institution for Science,
Carnegie Mellon University, the Chilean Participation Group, the French Participation Group, Harvard-Smithsonian Center for Astrophysics,
Instituto de Astrof\'isica de Canarias, The Johns Hopkins University,
Kavli Institute for the Physics and Mathematics of the Universe (IPMU) /
University of Tokyo, Lawrence Berkeley National Laboratory,
Leibniz Institut f\"ur Astrophysik Potsdam (AIP),
Max-Planck-Institut f\"ur Astronomie (MPIA Heidelberg),
Max-Planck-Institut f\"ur Astrophysik (MPA Garching),
Max-Planck-Institut f\"ur Extraterrestrische Physik (MPE),
National Astronomical Observatories of China, New Mexico State University,
New York University, University of Notre Dame,
Observat\'ario Nacional / MCTI, The Ohio State University,
Pennsylvania State University, Shanghai Astronomical Observatory,
United Kingdom Participation Group,
Universidad Nacional Aut\'onoma de M\'exico, University of Arizona,
University of Colorado Boulder, University of Oxford, University of Portsmouth,
University of Utah, University of Virginia, University of Washington, University of Wisconsin,
Vanderbilt University, and Yale University.

\bibliographystyle{model2-names}
\nocite{*}

\bibliography{vis}
\end{document}

%% file: sections/interaction.tex
Beyond the basic panning and zooming, further interactions with the sky map are carried out in three directions:
the ability to easily and interactively query the catalog from the MongoDB database, the ability to filter the objects in the map based on their properties, and the ability to apply different colormaps to the displayed objects.
\textit{Vizic} utilizes advanced web technologies, such as HTML5 and D3.js\footnote{https://d3js.org} \citep{d3}, as well as the framework provided by \texttt{ipywidget}\footnote{https://github.com/ipython/ipywidgets} to make such interactivity possible.
\subsubsection{Data Query}
\label{data query}
\textit{Vizic} provides two methods to interactively query the catalog data. The first method is carried out through the selection tool and the query button widget. The selection tool allows users to make lasso-like selections on a sky map, where the geospatial coordinates of the selection bound are automatically stored in the \texttt{AstroMap} object.
When the query button is clicked, \textit{Vizic} sends a request to the database, asking for all of the objects that resides within the selection bounding box. The MongoDB engine retrieves the requested objects using the geospatial index and returns them to the IPython kernel where \textit{Vizic} then reads the returned data using a \textit{pandas} DataFrame and assign that DataFrame to the \texttt{select\_data} attribute of the corresponding \texttt{GridLayer} object.

The second method is to directly click on the displayed astronomical sources. Since these objects are drawn using Scalable Vector Graphics (SVG) elements instead of a HTML canvas, each object can individually responds to mouse events.
When such a SVG element catches a ``click'' event, it queries the database for the particular object it represents.
The returned data is first parsed into a \textit{pandas} Series object and then assigned to the \texttt{object\_catalog} attribute of the \texttt{GridLayer} object.
To continuously observe the data returned while clicking through different objects, a user can create a \texttt{PopupVis} widget for the rendered \texttt{GridLayer} widget. The \texttt{PopupVis} widget displays the data in a HTML table and constantly checks for updates from the \texttt{object\_catalog} attribute.

The selection tool mentioned and the \texttt{PopupVis} widget can be found in Figure \ref{fig:overall}, which shows a customized GUI built using the widgets provided by \textit{Vizic}.
Once data is retrieved from the MongoDB, it can be further processed from other notebook cells, providing a continuous analysis framework.

\subsubsection{Data Filter}
\label{filter}
\begin{figure}[h]
\centering
\includegraphics[width=0.43\textwidth]{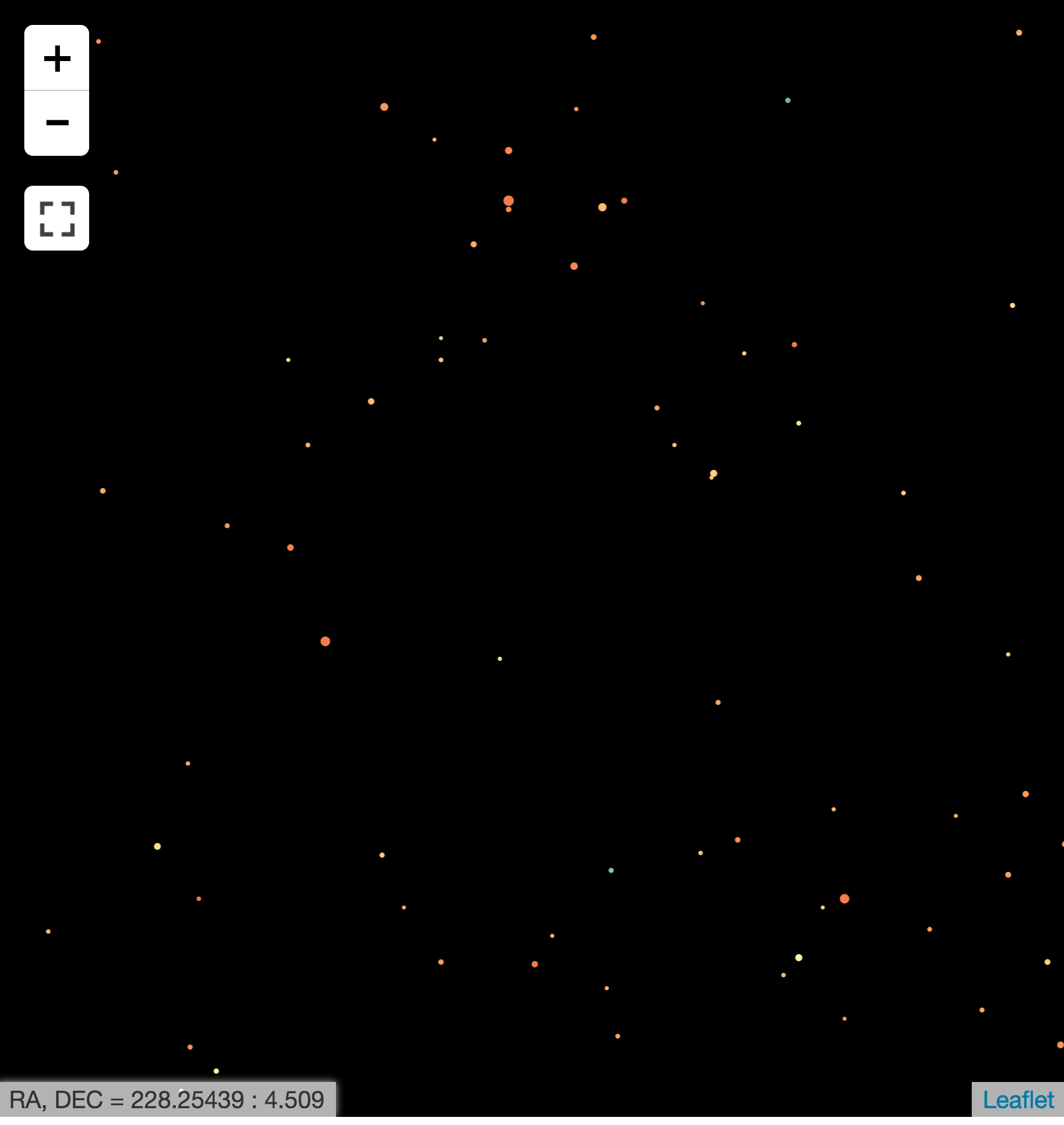}
\caption{The same area as shown in Figure \ref{fig:overall}, but only with object that has an $I$ magnitude \citep{sdss_filter} between $17.955$ and $27.249$.}
\label{fig:filtered}
\end{figure}
\textit{Vizic} also allows the users to filter displayed objects by their properties. In Jupyter notebooks, the filtering is controlled by a range slider bar.
The slider bar and another dropdown menu are wrapped into a single control widget, \texttt{FilterWidget}.
The dropdown menu in this control widget is responsible for switching the property field used to filter the objects. If a new property field is selected from the dropdown menu, the slider bar will updates itself with the new maximum value and minimum value for that particular field.
The property fields used to filter the object as well as the maximum and minimum values for each field are automatically determined and stored into the meta document\footnote{A document with `\_id' equals `meta'} in the catalog collection during the data ingestion process.

Every time the selected range on the slider bar changes, the \texttt{GridLayer} object validates the new range, updates the record and pushes the change to the front-end. As soon as the front-end widget receive the updates, it uses D3.js to select all objects that are outside the specified range and hides them. If a new property field is selected, \textit{Vizic} resets the filtering parameters and revalidates each object.

\begin{figure}[h]
\centering
\includegraphics[width=0.45\textwidth]{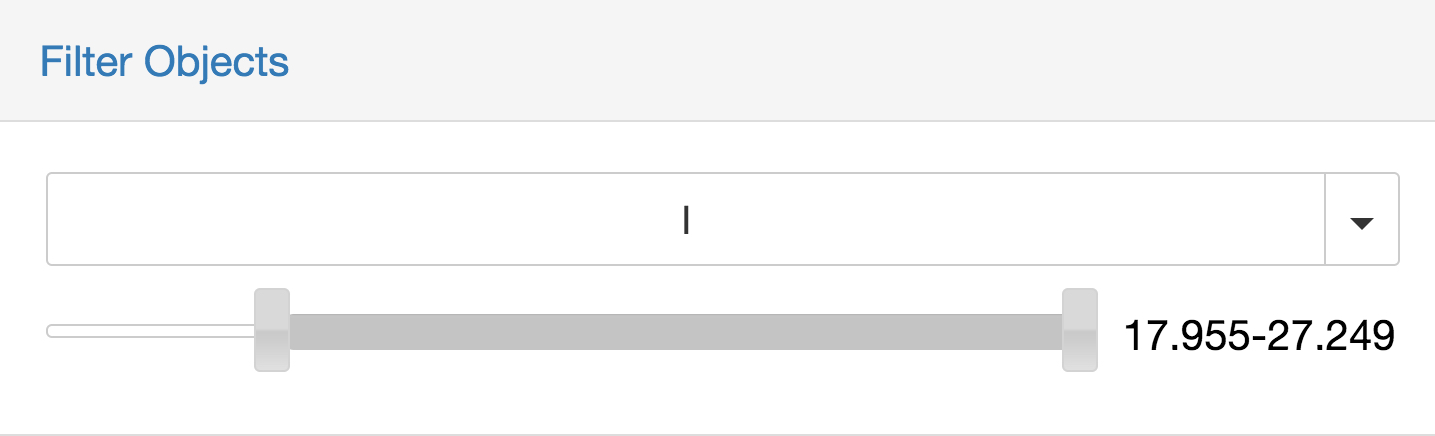}
\caption{A screenshot showing the FilterWidget}
\label{fig:slider}
\end{figure}

The objects filtering feature can be beneficial in many scenarios. For instance, we can easily determine outliers in the catalog by observing changes on the sky map while moving the slider bar from one side to another.
If the redshifts are provided, we can also visually inspect how cosmic structure evolves by shifting the selected range on the slide bar.
In Figure \ref{fig:filtered}, we shows the same area of the sky displayed by Figure \ref{fig:overall}, but only displaying the objects with an $I$ band magnitude between $17.955$ and $27.249$. Figure \ref{fig:slider} is a screenshot of the \texttt{FilterWidget} used.

\subsubsection{Color Maps}
\label{color maps}
\begin{figure}
\centering
\includegraphics[width=0.45\textwidth]{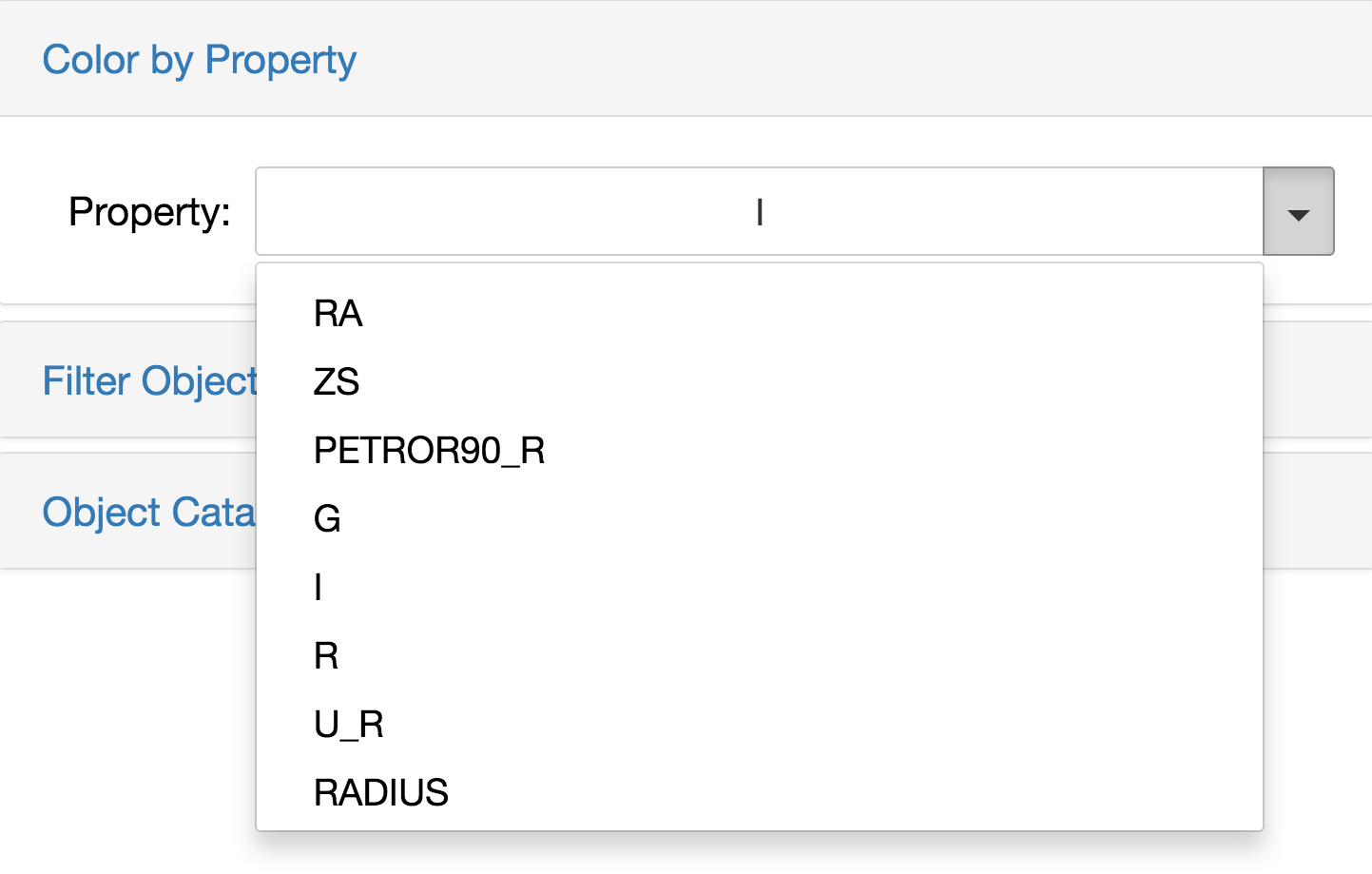}
\caption{A dropdown menu example showing properties that can be used for applying colormaps}
\label{fig:cdrop}
\end{figure}

Color mapping is another powerful feature offered by \textit{Vizic} for visual inspections of a given dataset. The default colormap used by \textit{Vizic} is the spectral scheme. Any of the property field available in the filtering widget could be selected for color mapping. The \texttt{CFDropdown} control widget is a dropdown menu (see Figure \ref{fig:cdrop}) that serves as an interface for switching the property field used for color mapping.
When the custom color mapping mode is enabled, the front-end JavaScript maps the value of the selected property from each object in the catalog to a range of $[0, 1]$ in a continuous and linear scale. Using the scaled value, \textit{Vizic} selects the matching color from the colormap and assigns the color value to the SVG element representing the particular object.
Beside the spectral scheme, \textit{Vizic} provides a variety of additional colormaps defined in D3.js\footnote{https://github.com/d3/d3-scale-chromatic}. The user can change the colormap applied using the control widget, \texttt{ColorMap}. As a new colormap is chosen, the colors for the SVG elements will be reassigned based on the new color space. However, if the property field being used is changed, the scaled value of each object is subject to recalculation.

%% file: sections/advanced.tex
\begin{figure*}[h!]
\centering
\subfloat[Voronoi diagram layer]{\includegraphics[width=0.4\textwidth]{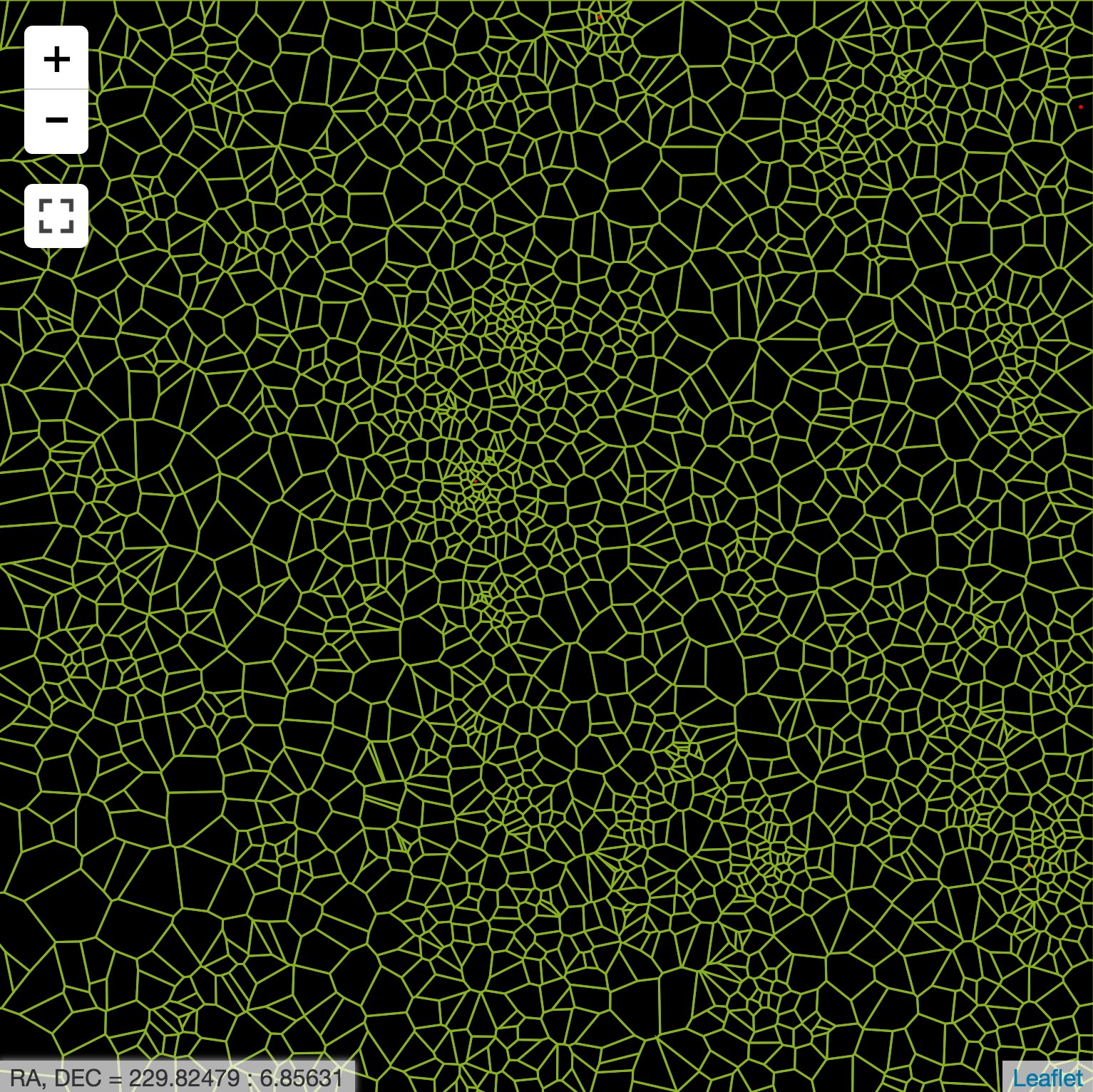}\label{fig:v_1}}
\quad
\subfloat[Minimum spanning tree layer]{\includegraphics[width=0.4\textwidth]{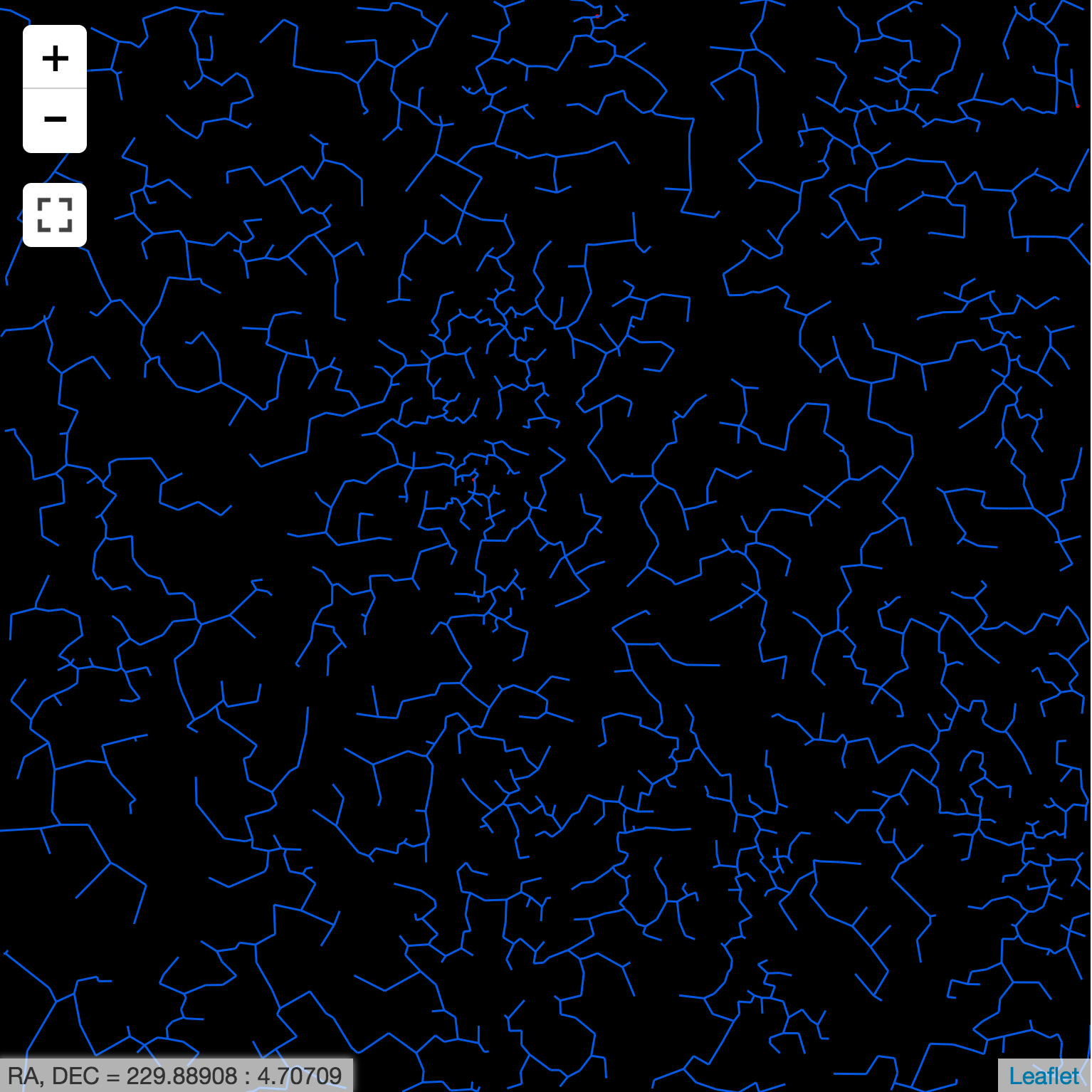}\label{fig:m_2}}
\quad
\subfloat[Delaunay triangulation layer]{\includegraphics[width=0.4\textwidth]{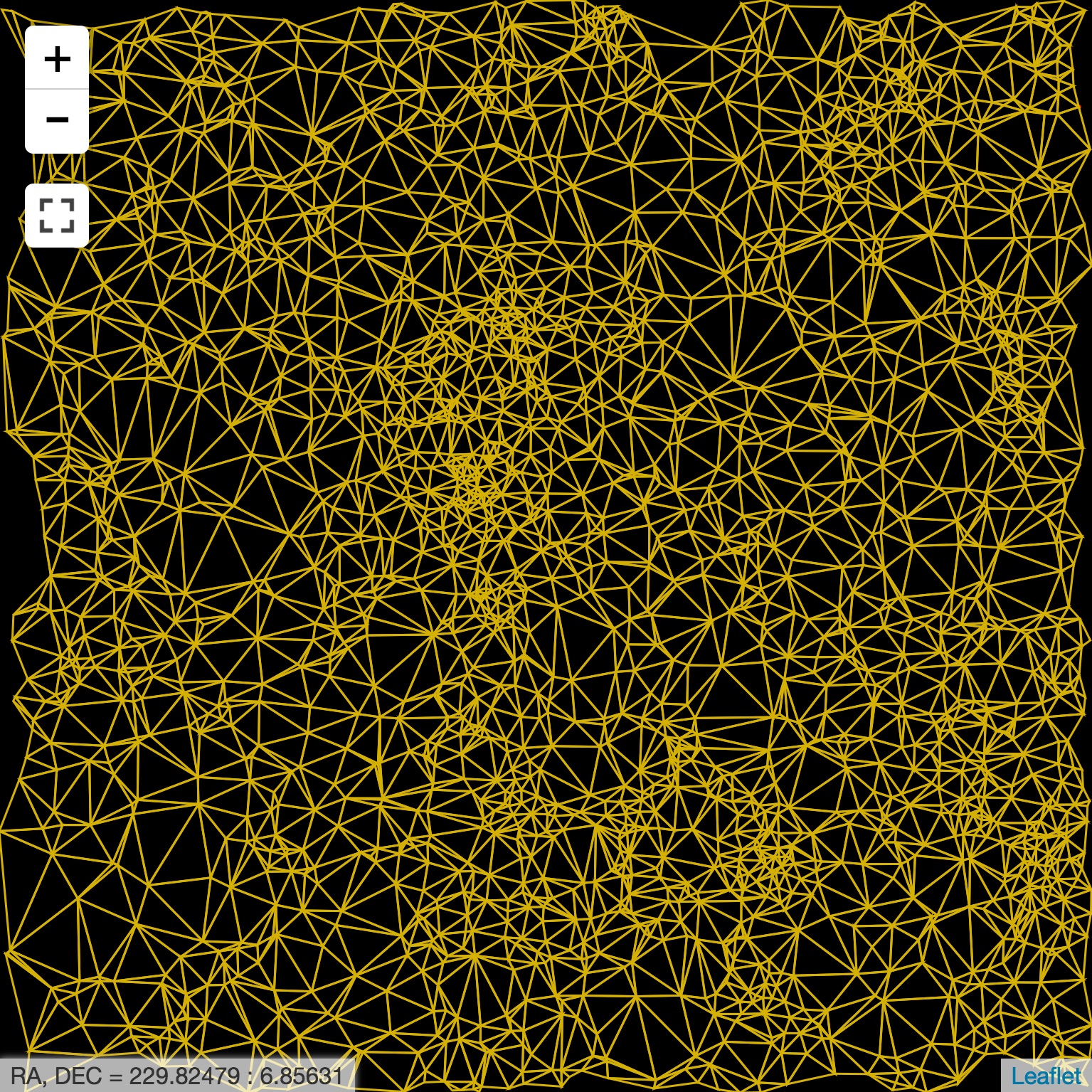}\label{fig:d_3}}
\quad
\subfloat[HEALPix layer. This overlay is created with $nside=1024$ and zoomed in to level 4 for a clear view of the content.]{\includegraphics[width=0.4\textwidth]{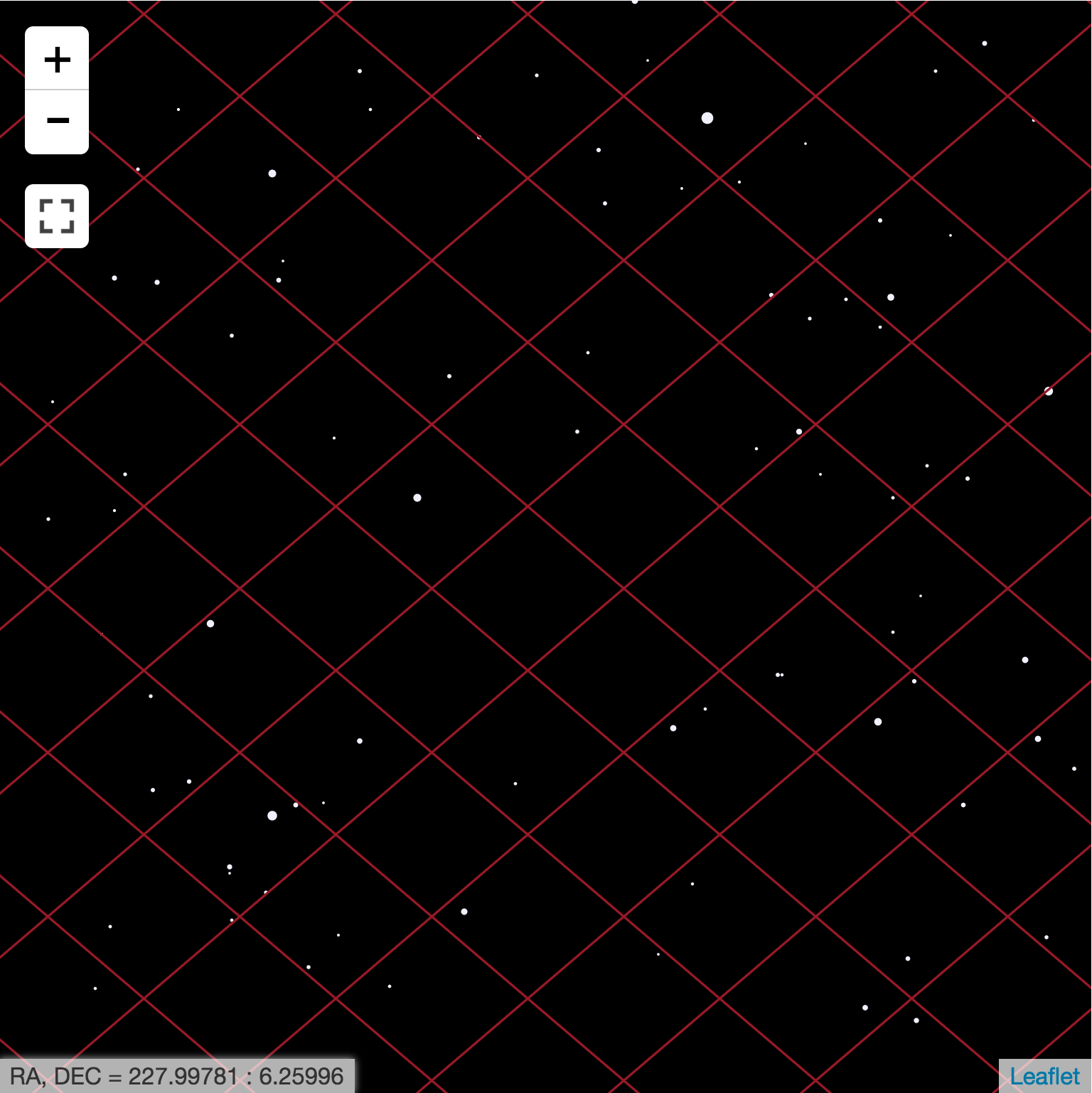}\label{fig:h_4}}
\caption{Custom overlays appended on the same area of the sky, they can be over plotted on top of one another.}
\end{figure*}
\textit{Vizic} allows custom overlays to be appended on top of the tile-based layer.
\textit{Vizic} provides four pre-defined overlay layers, a Voronoi diagram \citep{Voronoi1908} layer, a Delaunay triangulation \citep{Delaunay} layer, a minimum spanning tree (MST) \citep{boruuvka1926, Kruskal1956} layer and a HEALPix\footnote{http://healpix.sourceforge.net} \citep{healpix} grid layer.
HEALPix stands for \textbf{H}ierarchical \textbf{E}qual \textbf{A}rea iso\textbf{L}atitude \textbf{Pix}elation.
These overlays are implemented through four different layer classes.
New layer widget can be added and removed from the map using the \texttt{add\_layer} function and the \texttt{remove\_layer} function of the \texttt{AstroMap} class respectively.
Once the overlay is added, it can scale and adapt itself to follow the interactive movements on the map. The rest of this subsection describes each overlay individually and briefly discusses its construction procedure.

\subsubsection{Voronoi Diagram Layer}
\label{voronoi}
A Voronoi diagram for $n$ given points (sites) in a plane divides up the plane into $n$ regions (Voronoi cells), such that each region contains only one site and for every point inside that region the associated site is always the closet site.
Each Voronoi cell is also a convex polygon. The shared side between any two polygons is the perpendicular bisector of the line segment connecting the two corresponding sites.
In astronomy, Voronoi diagram is commonly used in the study of cosmic structure \citep[e.g.,][]{voronoiLss, voronoiLss2, Weygaert2007}.
\textit{Vizic} generates the Voronoi diagram by treating astronomical sources as sites and computing the diagram based on their locations in the sky.

The Voronoi diagram overlay for a particular sky map can be created by initializing a \texttt{VoronoiLayer} object with the corresponding \texttt{GridLayer} object as the first argument.
Inside the \texttt{VoronoiLayer} initialization function, we first query the database for all objects on the given tiled catalog layer and convert their celestial coordinates into pixel coordinates on the screen following the projection mechanism described in Section \ref{basic}.
Then we compute the diagram for these objects and render the diagram by drawing the Voronoi cells using SVG polygons, which allows another level of customization such as color, width, etc.

\subsubsection{Delaunay Triangulation Layer}
We have also implemented a Delaunay triangulation overlay layer. The Delaunay triangulation for a set of points in a plane is a triangulation such that none of the points lies inside the circumcircles of the triangles in this triangulation.
A Delaunay triangulation can also be seen as the dual graph of the Voronoi diagram for the same set of points.
The Delaunay triangulation layer can be created using \texttt{DelaunayLayer} class. Similar to how we construct the Voronoi diagram overlay, we first project the given objects in a catalog onto the screen and calculate the Delaunay triangulation from the positions of these objects, finally we use SVG path elements to connect the vertices of each determined triangle.

\subsubsection{Minimum Spanning Tree Layer}
\label{mst}
In graph theory, a spanning tree of a connected and undirected graph is a tree that contains all the nodes in that graph.
If each edge in a given connected graph is assigned a weight, a MST of that graph is a spanning tree such that the total weight for all edges in this tree is the least.
In this overlay, we use astronomical objects as the nodes and assign each edge a weight equals to its length measured in degrees.

The minimum spanning tree is computed using a combination of packages in the SciPy Stack \citep{numpy, pandas, scipy, scikit-learn}. We first use \texttt{sklearn.neighbors} \citep{scikit-learn} to calculate the distance between each object and its nearest 20 (default value) neighbors and then determine the MST from all possible trees that can be created using the given edges \citep{scipy, Kruskal1956}.
The result is a compressed sparse matrix where the entry is zero if no edge exists between the two objects indexed by the row number and column number respectively.
By connecting the objects referred by each non-zero entries, we obtain the MST.

\begin{figure}[h]
  \centering
  \includegraphics[width=0.43\textwidth]{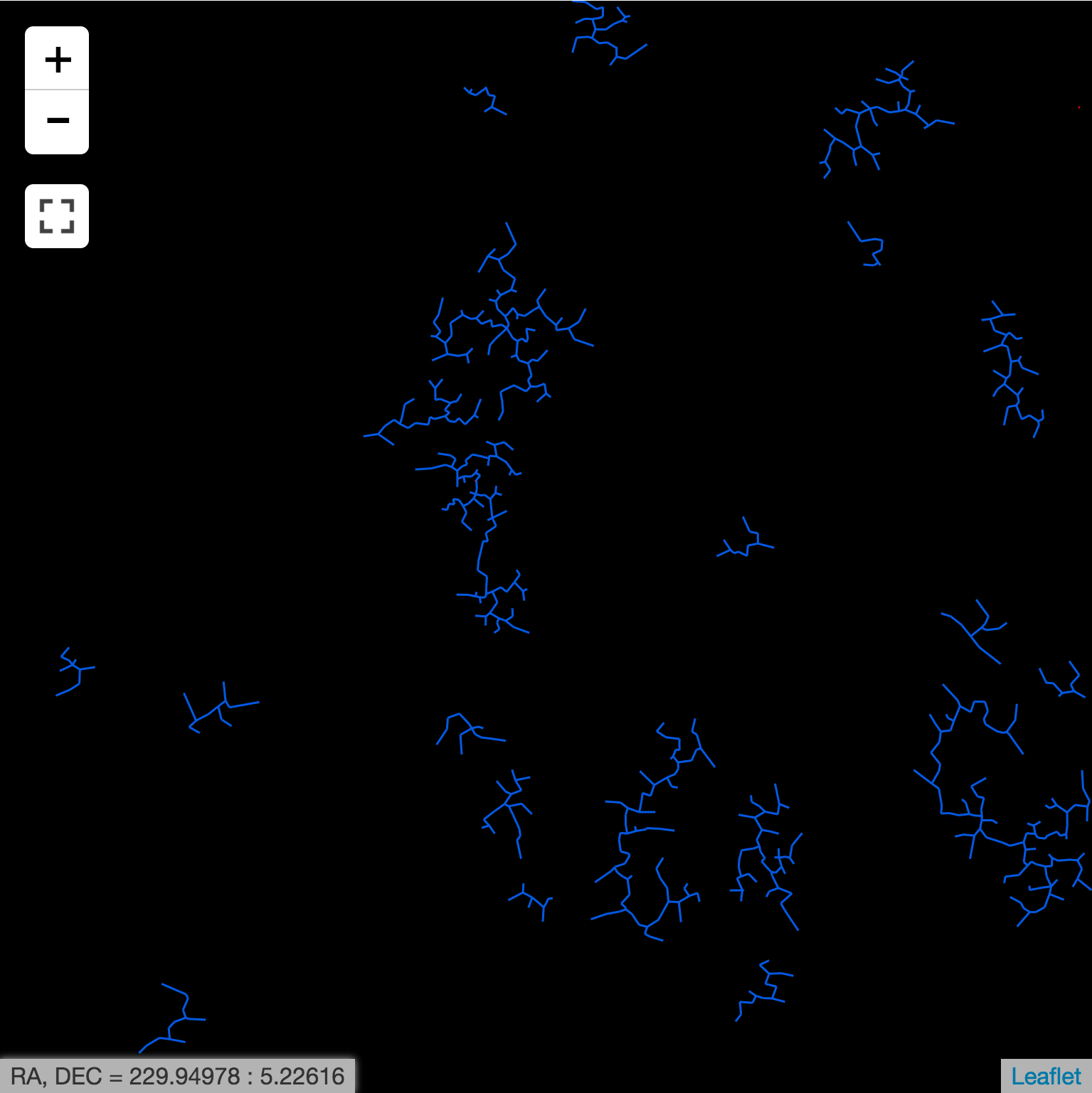}
  \caption{The tirmmed MST with $r_{max}=0.055^\circ$ (0.66 Mpc/h at z=0.2) and $n_{min}=8$}
  \label{fig:mst_trimmed}
\end{figure}

The minimum spanning tree overlay can be created by initiating a \texttt{MstLayer} object with the \texttt{GridLayer} object as the first argument.
Beyond visualizing the MST for a given set of objects on the sky map, we can also cut off the branches to select different structures \citep[e.g.,][]{Barrow1985,mst2}.
The user can prune the tree using the function \texttt{MstLayer.cut($r_{max}$, $n_{min}$)} with $r_{max}$ being the allowed maximum weight for saved edges and $n_{min}$ being the minimum number of edges that a saved sub branch must include.
As a result, only structures in a single branch with edges smaller than $r_{max}$ are kept.
Figure \ref{fig:mst_trimmed} shows a trimmed MST with a maximum weight of 0.055 degree or approximately 0.66 Mpc/h at a redshift (z) of 0.2 for groups with a minimum of eight members.

\subsubsection{HEALPix Grid Layer}
HEALPix is a method of pixelation that divides a spherical surface into quadrilaterals of equal area.
Each quadrilateral, in this case, is consider as a pixel. The lowest resolution of a HEALPix grid consist of 12 pixels and each higher resolution increase the total number pixels by four.
In our implementation, we use \textit{healpy}\footnote{https://github.com/healpy/healpy} to create the HEALPix grid, then project and connect the vertices of each pixel to draw the grid.
A demonstration of an appended HEALPix layer is shown in Figure \ref{fig:h_4}.

To display a HEALPix grid, a \texttt{HealpixLayer} object needs to be created first, with the \texttt{GridLayer} object of the target catalog layer as the first argument.
The resolution of the HEALPix grid can be specified by giving a new value to the keyword argument, \texttt{nside}, where the default is $1024$.
Like other overlay layers, a HEALPix grid overlay can be added or removed from the sky map using the \texttt{add\_layer} or \texttt{remove\_layer} function.
The advantage of using HEALPix grid is that we can then easily compute densities or other quantities per HEALPix pixel, which will improve the interaction with the datasets further allowing extra types of density based analysis.

%% file: sections/architecture.tex
\subsection{Extension Bundle}
\label{extension}
Technically speaking, \textit{Vizic} is an extension bundle to the Jupyter Notebook App. It comes with a notebook extension and a server extension.
The notebook extension is written in both Python and JavaScript. The Python code mainly consists of the APIs to create interactive widgets and a set of functions to interact with them.
The JavaScript section is the front-end interface that actually constructs the widgets in the browser, and also handles the interactions on the widgets by either sending state changing message to the kernel or responding directly at the front-end.
The Python code that runs in the kernel and its JavaScript counterpart that runs in the browser communicate with each other through the ``comms'' messaging API provided by the Jupyter Notebook.

Beyond the excellent platform offered by Jupyter Notebook, \textit{Vizic} also takes advantage of \textit{ipywidgets} and \textit{ipyleaflet}\footnote{https://github.com/ellisonbg/ipyleaflet}.
\textit{ipywidgets} offers a widget-model-view framework for building widgets within Jupyter notebooks.
All widgets in \textit{Vizic} are built based on the widget-model-view structure.
The widget-model-view structure uses IPython \textit{traitlets}\footnote{https://traitlets.readthedocs.io/en/stable/} to keep attributes or states synchronized at both front and back end and to trigger further actions according to state changes, which makes it possible to complete tasks both interactively using widgets and from the command line within the notebook cells.
IPython \textit{traitlets} is a framework that lets Python classes have attributes with type checking, dynamically calculated default values, and ‘on change’ callbacks. That way the user can build custom widgets to control the state of the sky maps and layers or to run and trigger additional scripts based on state changes. Such flexibility makes \textit{Vizic} even more powerful and extensible.

In addition, \textit{ipyleaflet} is a notebook extension package that integrates the \textit{Leaflet} library into the Jupyter Notebook framework environment.

The server extension is written also in Python using \textit{Tornado}\footnote{http://www.tornadoweb.org/}. This extension connects the kernel and the map widgets to the MongoDB database.
Whenever the front-end sky map asks for new data, the request will go through the server extension and return with the query results.
To enhance the performance, we used \textit{Motor}\footnote{https://motor.readthedocs.io}, an asynchronous Python driver for MongoDB, to access the database from the server.
As a result, multiple tiles on a map can be loaded concurrently, and multiple maps visualizing different catalogs can be displayed in one single notebook without a heavy performance cost.
\subsection{Slippy Map}
\begin{figure}[h]
  \centering
  \includegraphics[width=0.45\textwidth]{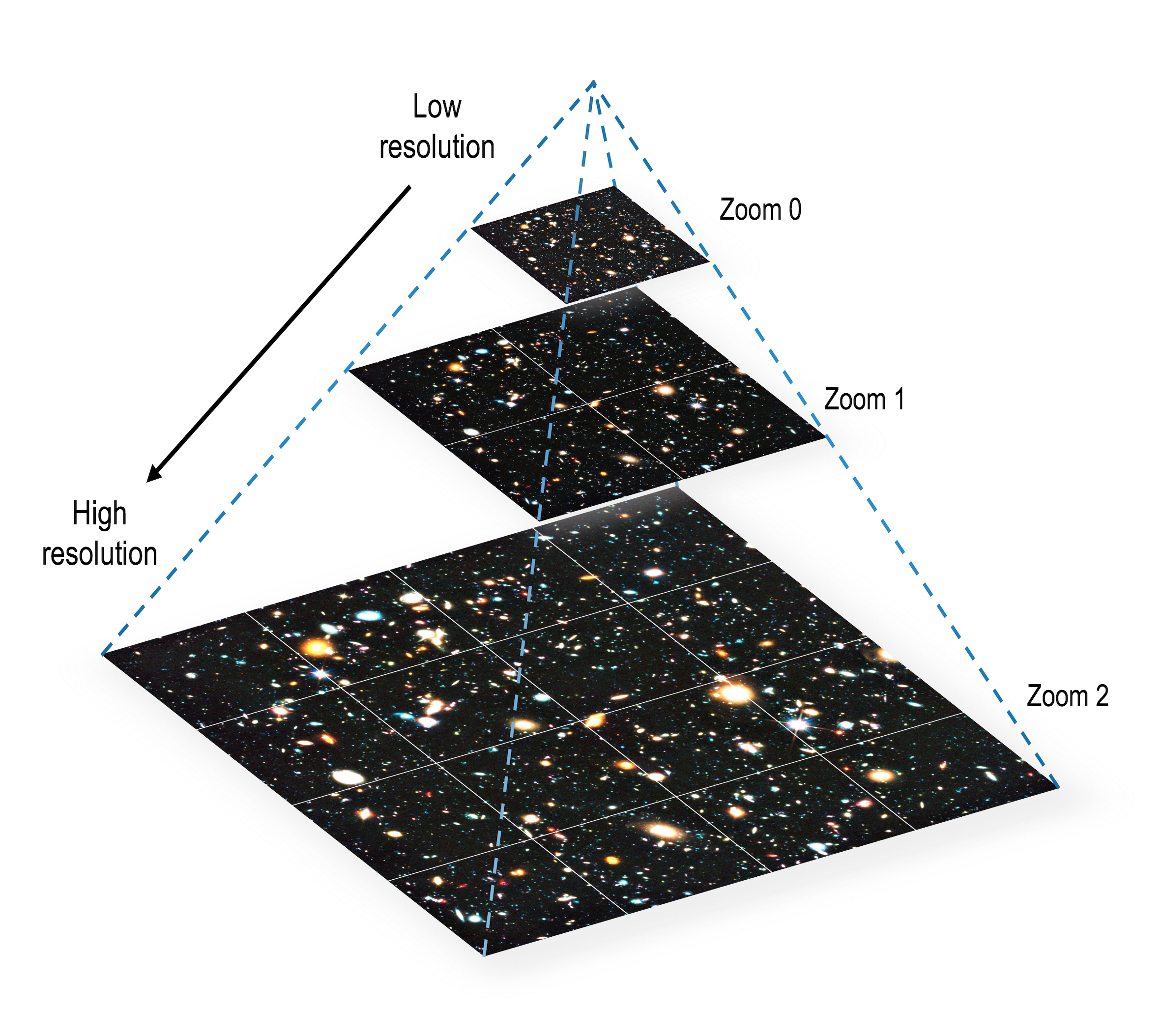}
  \captionsetup{width=0.45\textwidth}
  \caption{3D diagram explaining the pyramid like structure of the tiles system. Image Credit: NASA, ESA, H. Teplitz and M. Rafelski (IPAC/Caltech), A. Koekemoer (STScI), R. Windhorst (Arizona State University), and Z. Levay (STScI)}
  \label{fig:tilepyramid}
\end{figure}
\begin{figure}[h!]
  \centering
  \includegraphics[width=0.4\textwidth]{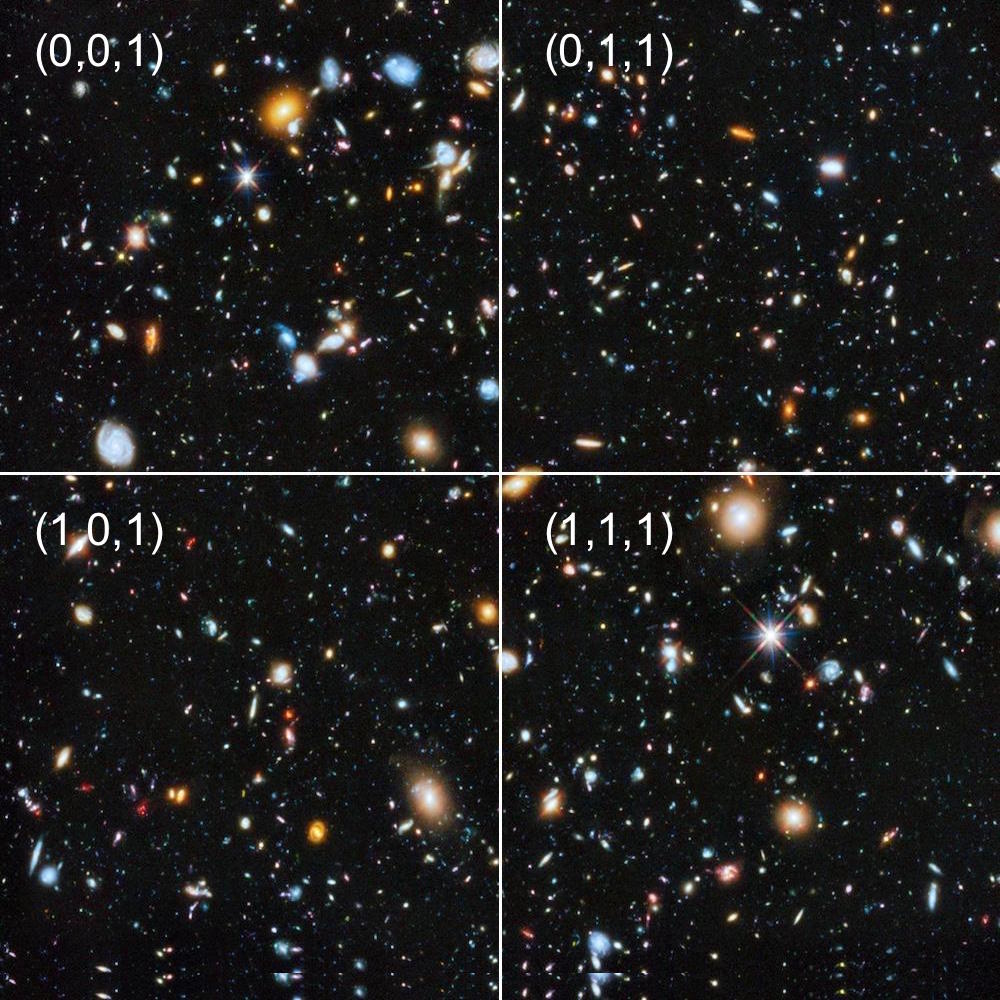}
  \captionsetup{width=0.45\textwidth}
  \caption{A figure showing the coordinate system used to locate the tiles. Image Credit: NASA, ESA, H. Teplitz and M. Rafelski (IPAC/Caltech), A. Koekemoer (STScI), R. Windhorst (Arizona State University), and Z. Levay (STScI)}
  \label{fig:coord}
\end{figure}

The main concept that stands behind the interactive maps created by \textit{Vizic} is the ``slippy map'' implementation.
The ``slippy map'' or tiled web map is the standard method of serving large maps on the web.
A well-known example adopting this approach is Google Maps. For a better illustration of the ``slippy map'' mechanism, we have constructed a 3-D diagram of its pyramid-like data structure using a Hubble image as an example (note that \textit{Vizic} does not display images), and with the diagram shown in Figure \ref{fig:tilepyramid}.
The images at different zoom levels are identical in terms of content but with different resolutions.
The higher the zoom level, the higher the resolution of the image. At each zoom level, the image is divided into many small square tiles, each with a size of $256 \times 256$ pixel.
The number of tiles at zoom level $n$ is be determined by
\begin{equation*}
N_{tile} = 2^{n} \times 2^{n}
\end{equation*}
\label{slippy}

Since the map window on our monitor is fixed, the number of small tiles that the window can display is relatively stable.
The JavaScript engine at the front-end only needs to request the tiles that can be displayed and disregard the rest for efficiency.
The initial idea behind this algorithm is that for an image that is much larger than the size of map window measured pixels, a big portion of this image is hidden at high zoom levels, which makes it very inefficient to load the entire image to the client side.
However, at lower zoom levels, many pixels are compressed into a single pixel on the display, therefore we are not able to perceive the information offered by the full-resolution image.
The ``slippy map'' implementation avoids the heavy loads by offering a pyramid-like multi-resolution structure, which in fact has been proved to be very efficient.
In order to correctly locate and request the tiles for \textit{Vizic}, each tile is also assigned an ID following the pattern $ (x, y, z) $, where $x$ and $y$ are row number and column number respectively and $z$ is the zoom level (see Figure \ref{fig:coord}).

Traditionally, in an interactive map, each tile is a PNG image, we have made a step forward by vectorizing these tiles before display.
We store each object in the catalog as a document in the MongoDB database and create the visual representations on the fly.
Before importing the catalog into the database, we split the data into $2^{n}*2^{n}$ groups following a similar grid as shown in Figure \ref{fig:coord} , where $n$ is the maximum zoom level with a default value of $8$, which can be changed at the catalog ingestion process.
Each group represents a tile at the maximum zoom level and objects in that group is assigned the tile ID of the group.
When a lower zoom level tile is requested, the ID for the requested tile is translated into a set of tile IDs at the maximum zoom level. This way we can further reduce the size and complexity of the data stored in the database.
The projection functions are shown below:
\begin{align*}
x_{max} &= xc*2^{z_{max}-z}-1\\
x_{min} &= (xc+1)*2^{z_{max}-z}\\
y_{max} &= yc*2^{z_{max}-z}-1\\
y_{min} &= (yc+1)*2^{z_{max}-z}
\end{align*}

$ x_{max}, x_{min}, y_{max}, y_{min} $ are the maximum and minimum values of $x$ and $y$ in the tile ID respectively.
$z_{max}$ is the maximum zoom level and $z$ is the zoom level of the requested tile.
At lower zoom levels, we reduce the load by leaving out objects that are too small to see (i.e., not resolved by the screen).

In particular, objects that have a semi-minor axis (or radius) smaller than one third of a pixel after being projected onto the screen will not be included in the data query. Here the number one third of a pixel is chosen because it provides extra tolerance for conversion offsets than one half of a pixel.
A shape that is smaller that one pixel either has a very low opacity or becomes invisible on the screen, so it is unnecessary to spend resource on those objects.
By projecting tile coordinates and leaving out invisible objects, we have successfully emulated the ``slippy map'' implementation in a vectorized way for catalog data.

The ``slippy map'' implementation is ideal for both displaying static images and displaying vectorized data, however the current scope of \textit{Vizic} is not to display static images. Even though, \textit{Vizic} can be extended to display both images and catalogs in the same manner.
\subsection{Custom Overlays}
\label{overlay_mech}
\begin{figure}[h]
  \centering
  \includegraphics[width=0.45\textwidth]{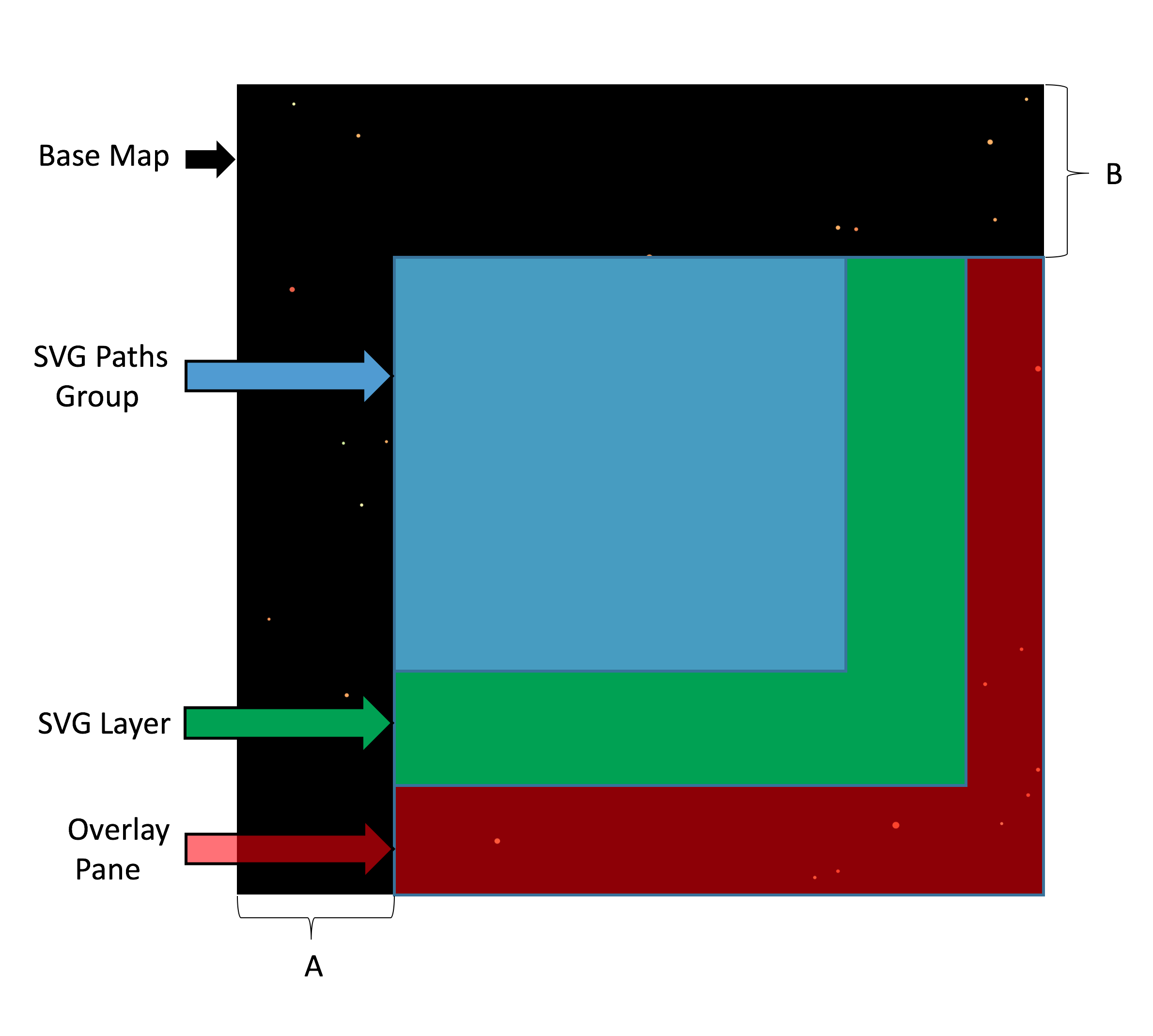}
  \captionsetup{width=0.45\textwidth}
  \caption{A diagram displays the layer structure of the mapping system. A is the offset in longitude direction. B is the offset in latitude direction.}
  \label{fig:layerdiagram}
\end{figure}
As described in Section \ref{adv}, \textit{Vizic} allows custom overlays to be appended on top of the visualized catalogs.
We make this possible by extending the \texttt{Layer} class in \textit{Leaflet} and providing specific drawing constructions based on the type of geometric shape contained in the overlay.
Since it is unpractical to redraw the overlay using canvas element every time the viewport changes (e.g., zoom and pan), we instead create the overlay with a collection of SVG elements.
SVG elements are vector-based, therefore we only need to create them once and resize them as the map size changes.
Because the \textit{Leaflet} map is a multi-layered system (see Figure \ref{fig:layerdiagram}), to accommodate the offsets generated by rescaling of the overlay when zoom level changes, we put these SVG elements into a group element, and scale and shift them together.
The offsets are determined as follows:
\begin{align*}
A &= lon_{base} - lon_{op}\\
B &= lat_{base} - lat_{op}
\end{align*}

$A$ and $B$ are the offsets in the longitude and latitude directions, $(lon_{base},lat_{base})$ is the coordinate for the base map and  $(lon_{op},lat_{op})$ is the coordinate for the overlay pane. In \textit{Leaflet}, panes are used to control the ordering of layers on the map. Overlay pane is the default pane for all the vector overlay layers.

\subsection{MongoDB}
\label{mongodb}
\textit{Vizic} uses MongoDB databases to store and serve the catalogs.
MongoDB is a document-oriented database, which stores rows from a traditional RDBMS database into documents to enable faster query and better scalability.
Considering astronomical catalogs are simply collections of observed objects without complex relations between each columns when it comes to visualizing them, and since \textit{Vizic} only asks for objects and their properties from a catalog using their positions in the sky, MongoDB is a better choice comparing to a traditional RDBMS database.

\textit{Vizic} stores the data for each object in a document, each collection of documents is a catalog and each database can contain many collections.
Every document is indexed by its tile ID and the size of the object measured in arcsecond to enhance query performance.
\textit{Vizic} also takes advantage of MongoDB's internal geospatial index to efficiently retrieve data from the selection tool.

Another important feature of MongoDB is that it allows flexible data structures within a document.
In other words, different documents within one collection can have different fields, which is extremely useful for visualizing a composite catalog where the objects come from multiple surveys and each survey provide their own measurements.
Using a relational database, such tasks would be more difficult to accomplish.